\documentclass{article}
\usepackage[utf8]{inputenc}
\usepackage[margin=1in]{geometry} 
\usepackage{amsmath}
\usepackage{tcolorbox}
\usepackage{amssymb}
\usepackage{amsthm} 
\usepackage{accents}
\usepackage{physics}
\usepackage{graphicx}
\usepackage{hyperref}
\usepackage[separate-uncertainty=true, per-mode=reciprocal]{siunitx}
\usepackage{xcolor}
\hypersetup{colorlinks, linkcolor={red!50!black}, citecolor={blue!50!black}, urlcolor={blue!80!black}}
\usepackage{enumerate}
\usepackage{mathrsfs}
\usepackage[makeroom]{cancel}
\newcommand{\nn}{\nonumber}

\DeclareMathOperator{\sign}{sgn}
\DeclareDocumentCommand\sgn{}{\trigbraces{\sign}}
\DeclareMathOperator{\EXPV}{E}
\DeclareDocumentCommand\Expv{}{\trigbraces{\EXPV}}
\DeclareMathOperator{\VARNC}{V}
\DeclareDocumentCommand\Varnc{}{\trigbraces{\VARNC}}
\DeclareMathOperator{\COVAR}{Cov}
\DeclareDocumentCommand\Covar{}{\trigbraces{\COVAR}}
\DeclareMathOperator{\PROBY}{P}
\DeclareDocumentCommand\Proby{}{\trigbraces{\PROBY}}
\DeclareMathOperator{\errorfunc}{erf}
\DeclareDocumentCommand\erf{}{\trigbraces{\errorfunc}}
\usepackage{bm}
\usepackage{mathtools}
\usepackage[sorting=none]{biblatex} 
\usepackage{authblk}

\title{Accuracy and capacity of Modern Hopfield networks with synaptic noise}

\author[1]{Sharba Bhattacharjee}
\author[2, 1]{Ivar Martin}

\affil[1]{Department of Physics, University of Chicago, Chicago, Illinois 60637, USA}
\affil[2]{Materials Science Division, Argonne National Laboratory, Lemont, Illinois 60439, USA}

\begin{document}

\maketitle
\begin{abstract} 
    We study the retrieval accuracy and capacity of modern Hopfield networks of with two-state (Ising) spins interacting via modified Hebbian $n$-spin interactions.  In particular, we consider systems where the interactions deviate from the Hebb rule through additive or multiplicative noise or through clipping or deleting interactions.  We find that the capacity scales as $N^{n-1}$ with the number of spins $N$ in all cases, but with a prefactor reduced compared to the  Hebbian case. For $n=2$ our results agree with the previously known results for the conventional $n = 2$ Hopfield network.
\end{abstract}
\section{Introduction}\label{intro}

The Hopfield network, proposed in 1982~\cite{hopfield_neural_1982}, is a system of binary degrees of freedom (usually termed Ising spins or binary neurons) with long-range interactions, which exhibits a form of behavior known as associative memory. Specifically, we can choose a set of spin configurations (hereby called patterns), and define the coupling constants for the spin-spin interactions such that those patterns are the locally stable  states of the system (with a spin configuration sufficiently close to any of the specified patterns  relaxing to that pattern). This behavior is similar to a human brain reconstructing a memory when provided with incomplete information, and so the neural networks such as the Hopfield network have been used to model this process ~\cite{amit_spin-glass_1985}. The coupling constants of the network are analogous to the strengths of the synaptic connections between neurons which encode biological memory; most commonly, the Hebb's learning rule is used to define them~\cite{hebb_organization_2005}. Hopfield networks have been physically realized in digital architecture~\cite{hendrich_scalable_1996} and confocal cavity QED systems~\cite{marsh_enhancing_2021}.

The number of distinct patterns $K$ that can be retrieved without significant error (network capacity) is determined by the number of spins $N$ in the system. From probabilistic considerations and Monte Carlo simulations, Hopfield estimated the capacity of a network of size $N$ to be $0.15N$~\cite{hopfield_neural_1982}. Using techniques from equilibrium statistical physics such as replica theory, Amit et al.~\cite{amit_information_1987} found that the overlap of a state of the system with the pattern closest to it at equilibrium can be nonzero for sufficiently few patterns (indicating small error), but jumps to zero when the number of patterns is approximately $0.138N$, exhibiting a first order phase transition. This critical number of patterns can thus be considered to be the capacity of the network as long as some retrieval errors were allowed. The above result assumed replica symmetry; later work which allowed for replica symmetry breaking~\cite{crisanti_saturation_1986}  showed a somewhat higher capacity of  $0.144N$. 
Using techniques from coding theory, McEliece et al.\ showed that if we require that for every pattern, every state within a Hamming distance of $\delta N$ from that pattern (for $\delta<1/2$) is retrieved perfectly after one step of synchronous update (with probability approaching $1$ as $N\to\infty$), then the capacity of the Hopfield network is $(1-2\delta)^2 N/4\ln(N)$~\cite{mceliece_capacity_1987}.

The capacity of the Hopfield network has also been estimated in models where the couplings are not defined by the Hebb rule. Abu-Mostafa et al. showed that the capacity for any system with two-spin interactions is bounded above by $N$ irrespective of how the couplings are defined~\cite{abu-mostafa_information_1985}. An important example of alternatively defined couplings is the model with ``quantized'' or ``clipped'' couplings, where the coupling constants all have the same magnitude and only the sign is determined by Hebb rule. McEliece et al.\ showed that the capacity for this model would be related to that of the Hebbian model by a factor of $2/\pi$~\cite{mceliece_capacity_1987}, which was verified by Sompolinsky using mean field theory when he found the capacities of a class of models with non-Hebbian couplings (including the clipped network as well as couplings with additive noise)~\cite{sompolinsky_theory_1987}. 

The above models can be naturally extended to involve simultaneous interactions between more than two spins. Spin-glass models with multi-spin interactions have been studied before in refs.~\cite{kirkpatrick_p_1987, gardner_spin_1985, crisanti_sphericalp-spin_1992, crisanti_sphericalp-spin_1993}, as well as in the context of error correcting codes in ref.~\cite{sourlas_spin-glass_1989}. Bovier and Niederhauser~\cite{bovier_spin-glass_2001} and Baldi and Venkatesh~\cite{baldi_number_1987} had considered generalizations of the Hopfield network with $n$-spin interactions and showed that the memory capacity of such systems scales with the system size $N$ as $N^{n-1}$. Recently, Krotov and Hopfield generalized the Hopfield network to include a more general class of Hamiltonians, giving rise to networks with much higher capacities, called dense associative memory models or modern Hopfield networks~\cite{krotov_dense_2016}. Krotov and Hopfield had estimated the capacity of such models~\cite{krotov_dense_2016}, and it was extended to include more general notions of capacity by Bao et al.~\cite{bao_capacity_2022}, which accounted for finite basins of attraction around the patterns. Ramsauer et al.~\cite{ramsauer_hopfield_2021} introduced a modern Hopfield network whose capacity scales exponentially as system size (as shown for more general ensembles of patterns in~\cite{lucibello_exponential_2024}), and showed this network to be equivalent to the attention mechanism used in transformer models, a kind of neural network~\cite{vaswani_attention_2023}. Such models have been analyzed in the context of Boltzmann machines~\cite{ota_attention_2023}, utilized to perform tasks such as in-context denoising~\cite{smart_-context_2025} and storing a sequence of patterns~\cite{chaudhry_long_2023}, and refinements to their dynamics have also been proposed to further improve their memory capacity~\cite{wu_uniform_2024, hu_provably_2024} as well as to propose novel architectures such as energy transformers~\cite{hoover_energy_2023}. A system of Ising spins with $n$-spin interactions is one such class of dense associative memory ($n = 2$ being the conventional Hopfield network). Agliari et al.\ studied the capacity of modern Hopfield networks with multi-spin interactions in presence of additive noise in the coupling constants (e.g.,  resulting from errors when learning or storing the Hebbian interaction coefficients)~\cite{agliari_tolerance_2020}. 
They found that the capacity of the network still scales with the system size as $N^{n-1}$ in all cases, but with a prefactor that depends on the particular modification.

In this paper, we consider modern Hopfield networks with $n$-spin interactions, where the interaction coefficients deviate from the Hebb rule in a more general way. In particular, we consider models with additive and multiplicative noise in the interactions and models where some of the interactions have been randomly deleted or where the interaction coefficients have been clipped to all have the same magnitude and  only differ in sign. We find that for the case of $n=2$, our results agree with those found previously for the non-Hebbian Hopfield network by McEliece and Sompolinsky. For general $n$, we find that the capacity scales with the system size $N$ as $N^{n-1}$  in the presence of all these noise channels, significantly generalizing the result of ref.~\cite{agliari_tolerance_2020}.


A complementary perspective to the memory storage/retrieval robustness described above is the one of the efficiency in terms of memory and computational requirements of a network. Storage of clipped  synaptic weights  requires only single bit per bond. Similarly, dropping a bond completely further reduces the storage requirements.  Binary weights also allow to perform the retrieval computation faster than general weights. Therefore, the networks with clipped and partially dropped couplings can potentially outperform the ``perfect" networks for the same memory storage tasks.  We demonstrate indeed that this is the case. 

We note that more sophisticated schemes exist for deleting less important couplings based on their information content, as a way to maximize the storage efficiency ~\cite{lecun_optimal_1989}. We do not consider this approach here; all interactions are modified randomly and  independently with the same probability.

The rest of the paper is organized as follows. 
In section~\ref{description_n}, we define the model of dense associative memory considered here. In section~\ref{retrieval}, we provide the definitions we have used for the capacity, following refs.~\cite{mceliece_capacity_1987} and ~\cite{bao_capacity_2022}. In section~\ref{dense}, we estimate the storage capacity of modern Hopfield networks using statistical techniques. We find the capacity for networks with Hebbian interactions as well as those with noise or with clipped and/or deleted interactions, and  compare their storage efficiencies. In section~\ref{numerics}, we numerically simulate modern Hopfield networks and compare the results to our analytical estimates from the previous section. 
In section~\ref{outlook}, we summarize our findings and mention possible applications and outlook.

\section{Definition of the modern Hopfield network with synaptic noise}\label{description_n}
\begin{figure}
    \centering
    \includegraphics[width=0.5\linewidth]{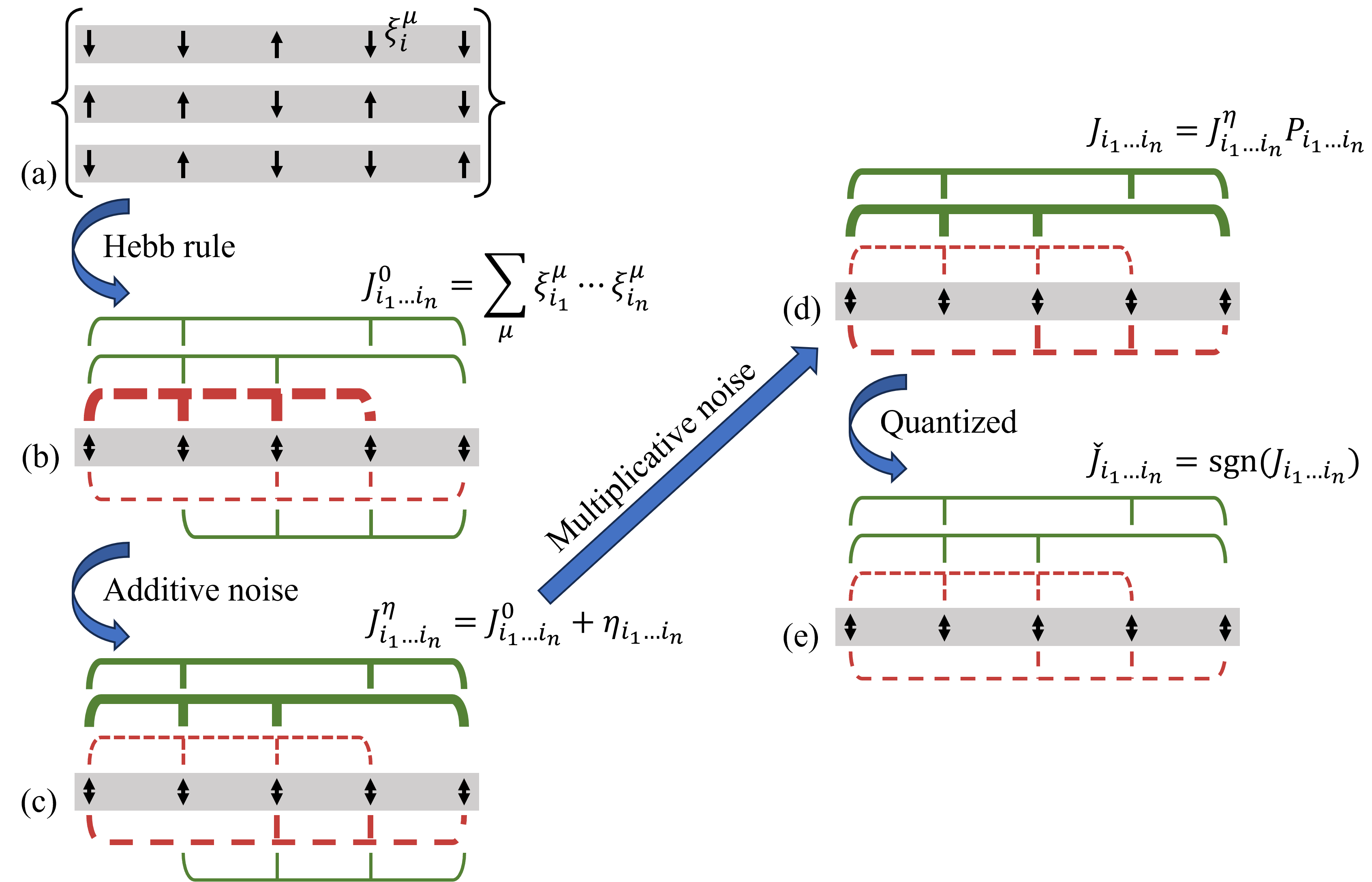}
    \caption{Schematic description of a modern Hopfield network with $N=5$ spins, $K=3$ patterns, and $n=4$ spin interactions. (a) The set of patterns $\qty{\vb* \xi^\mu}$ used to define the interaction coefficients, with components $\xi^\mu_i\in\qty{\pm 1}$ (shown as upward and downward arrows). (b) The $n$-spin interaction coefficients $J^0_{i_1\ldots i_n}$ defined by the generalized Hebb rule. In this figure, each interaction connects $n=4$ spins, shown by a branching line segment connecting the spins. The thickness of the line segment indicates the strength $\abs{J^0_{i_1\ldots i_n}}$ of the interaction. Positive interaction coefficients are shown using solid green lines, and negative ones using dashed red lines. (c) The interactions with additive noise $\eta_{i_1\ldots i_n}$; in this example, they are independent Gaussian $N(\mu=0, \sigma^2=4)$ random variables. (d) The interactions with additive as well as multiplicative noise $P_{i_1\ldots i_n}$. In this example the multiplicative factors are Bernoulli random variables $B(1,0.8)$, which effectively ``deletes'' a fifth of the interactions. (e) Description of the alternate model considered in this paper, where the interactions are clipped to all have the same strength and differ only in sign.}
    \label{fig:quarticscheme}
\end{figure}

Consider a system of $N$ Ising spins used to store $K$ patterns $\vb*\xi^\mu$, where we suppose that $N, K \gg 1$. The pattern components $\xi^\mu_i$ have been chosen randomly and are independent and identically distributed Rademacher random variables, with probability distribution $\Proby(\xi^\mu_i=1)=\Proby(\xi^\mu_i=1)=1/2$, and moments $\Expv(\xi^\mu_i)=0$, $\Expv({\xi^\mu_i}^2)=1$. 
Given set of memory patterns, we can define the Hamiltonian for an arbitrary state  $\vb*\sigma = (\sigma_1, ..., \sigma_N)\in\qty{1,-1}^N$, 
\begin{align}
    H^0_n(\vb*\sigma) &= -\frac{1}{N^{n-1}}\sum_{1\le i_1<\ldots<i_n\le N} \sum_{\mu=1}^K \xi^\mu_{i_1}\cdots\xi^\mu_{i_n} \sigma_{i_1}\cdots\sigma_{i_n} = -\frac{1}{n! N^{n-1}}\sum_{i_1\neq\ldots\neq i_n} \sum_{\mu=1}^K \xi^\mu_{i_1}\cdots\xi^\mu_{i_n} \sigma_{i_1}\cdots\sigma_{i_n},\label{nham}
\end{align}
which is the model with $n$-spin interaction studied by Agliari et al.~\cite{agliari_tolerance_2020}. For $n>2$, this is a dense associative memory model as defined by Krotov~\cite{krotov_dense_2016}, which is made apparent by the fact that the above Hamiltonian can be written as a sum over $n$-th order polynomial functions of the overlaps $m_\mu = \sum_{i=1}^N \xi^\mu_i \sigma_i/N$ of the system with the patterns.  In particular, as shown in appendix~\ref{funfact}, in the limit of large $N$, we have 
\begin{align}
    H^0_n(\vb*\sigma) &= -N \sum_{\mu=1}^K \sum_{k=0}^{\lfloor n/2 \rfloor} (-1)^k \frac{m_\mu^{n-2k}}{k!(n-2k)!(2N)^k},
\end{align}
where the lower order corrections (corresponding to $k>0$) appear because we have excluded self-interactions which effectively reduce the order of interaction. We note that for an arbitrary state $\vb*\sigma$ and pattern $\vb*\xi^\mu$, the overlap is of the order $m_\mu\sim 1/\sqrt{N}$, in which case every term in the above expression is of order $N^{1-n/2}$; however, if the state is close to a particular pattern (say $\vb*\xi^1$), then $m_1\sim 1$, and the terms in the sum over $k$ have orders of $N$ and lower.

The Hamiltonian $H^0_n$ in eq.~\eqref{nham} describes an Ising spin system system with $n$-spin interactions, mediated by the interaction constants $J^0_{i_1\ldots i_n} = \sum_\mu \xi^\mu_{i_1}\cdots \xi^\mu_{i_n}$; we assume that $N\gg n$. This is the extension of the Hebb rule to higher order interactions. We can now generalize the model to introduce some synaptic noise in the interaction constants. Specifically, we consider the situation where the Hamiltonian is
\begin{align}
    H_n(\vb*\sigma) = -\frac{1}{N^{n-1}}\sum_{1\le i_1<\ldots<i_n\le N} J_{i_1\ldots i_n} \sigma_{i_1}\cdots\sigma_{i_n},
\end{align}
which corresponds to $n$-body interactions mediated by the interaction coefficients
\begin{align}
J_{i_1\ldots i_n} = \qty(\sum_\mu \xi^\mu_{i_1}\cdots \xi^\mu_{i_n} + \eta_{i_1\ldots i_n})P_{i_1\ldots i_n},
\end{align}
where the random variables $\eta_{i_1\ldots i_n}$ and $P_{i_1\ldots i_n}$ correspond to additive and multiplicative noise respectively. We assume that  all of the random variables are independent. The variables $\eta_{i_1\ldots i_n}$ are taken to have identical probability distributions, and thus have the same moments $\Expv(\eta_{i_1\ldots i_n})=\mu_\eta$ and $\Expv(\eta_{i_1\ldots i_n}^2)=\mu_{\eta^2}$. In particular, we can consider the case where $\eta_{i_1\ldots i_n} \sim N(0, \mu_{\eta^2})$, corresponding to Gaussian additive noise. Such additive noise changes the interaction strengths and may also flip the sign of some weaker interactions. Similarly, the variables $P_{i_1\ldots i_n}$ corresponding to multiplicative noise also have identical probability distributions, with moments $\Expv(P_{i_1\ldots i_n})=\mu_P$ (which we will assume to be positive) and $\Expv(P_{i_1\ldots i_n}^2)=\mu_{P^2}$. Such multiplicative noise can also affect the interaction strengths, and it can also randomly delete some interactions (if $P_{i_1\ldots i_n}$ is zero) or flip their sign (if $P_{i_1\ldots i_n}$ is negative). 

The mapping $\vb T_n$ corresponding to one step of synchronous update of the system, according  to the Hamiltonian with interaction constants defined above, is
\begin{align}
    (\vb T_n\vb*\sigma)_i &= \sgn(\sum_{\substack{1\le i_2\le\cdots\le i_n\le N,\\i_a\neq i\forall a}} J_{ii_2\ldots i_n}\sigma_{i_2}\cdots\sigma_{i_n}) \nn\\
    &= \sgn(\sum_{\mu,\qty{i_a}} \xi^\mu_i \xi^\mu_{i_2}\cdots \xi^\mu_{i_n}\sigma_{i_2}\cdots\sigma_{i_n} P_{i_1\ldots i_n} + \sum_{\qty{i_a}} \sigma_{i_2}\cdots\sigma_{i_n} \eta_{i_1\ldots i_n} P_{i_1\ldots i_n}) \label{update}
\end{align}

\section{Retrieval and capacity}\label{retrieval}
The Hopfield network is a model for associative memory, where the patterns $\vb*\xi^\mu$ are the stable states satisfying $\vb T(\vb* \xi^\mu)=\vb*\xi^\mu$ for all $\mu$, and states that are close to a pattern (in Hamming distance) are mapped to the corresponding pattern (``retrieved'') under the Hopfield update $\vb T$. To quantify this, we follow the work of McEliece et al.~\cite{mceliece_capacity_1987} and Bao et al.~\cite{bao_capacity_2022}, and consider the following setup:
\begin{enumerate}
    \item Start with an arbitrary stored memory pattern, say $\vb*\xi^1$.
    \item Perturb the pattern by flipping each of its component spins independently with probability $\delta$ ($0\le\delta< 1/2$) to obtain the state $\tilde{\vb*\xi}^1$, where $\tilde \xi^1_i = s_i \xi_i^1$ for i.i.d.\ random variables $s_i$ taking the value $1$ ($-1$) with probability $1-\delta$ ($\delta$). This is equivalent to retaining $1-2\delta$ of the original pattern components and assigning the rest randomly to be $\pm 1$ with equal probability. $\tilde{\vb*\xi}^\mu$ has a Hamming distance of $N\delta$ from $\vb*\xi^\mu$ on average, which can be used to define the basin of attraction around each pattern that we would want to be retrieved. $\delta=0$ corresponds to the case where we only care if the patterns themselves are stable.
    \item Perform one step of the Hopfield update on $\tilde{\vb*\xi}^1$ to obtain the state $\vb T_n\tilde{\vb*\xi}^1$, where
    \begin{align}
        (\vb T_n\tilde{\vb*\xi}^1)_i = \sgn(\sum_{i_2\neq\cdots\neq i_n\neq i} J_{i i_2\ldots i_n}\xi^1_{i_2}\ldots \xi^1_{i_n} s_{i_2}\ldots s_{i_n}).
    \end{align}
    \item Find the overlap $m = \sum_i \xi^1_i (\vb T_n \tilde{\vb*\xi}^1)_i/N=\sum_i \sgn(\sum_{i_2\neq\cdots\neq i_n\neq i} J_{ii_2\ldots i_n}\xi^1_i \xi^1_{i_2}\cdots \xi^1_{i_n} s_{i_2}\cdots s_{i_n})/N$ of the obtained state with the original pattern $\vb*\xi^1$ to quantify how close it is to the original pattern. $m=1$ corresponds to perfect retrieval, $\vb T_n \tilde{\vb* \xi}^1=\vb*\xi^1$.
\end{enumerate}
The overlap $m$ defined as above is itself a random variable, where $m=1$ corresponds to perfect retrieval. Therefore, we can use $m$ to quantify how accurately the patterns are retrieved. One measure of retrieval quality is the expected overlap, $\Expv(m)$. This is a convenient measure, as it is easy to estimate using the specified probability distributions of $\xi^\mu_i$, $\eta_{i_1\ldots i_n}$, $P_{i_1\ldots i_n}$, and $s_i$, and invoking central limit theorem where appropriate. Defining $m_i = \sgn(\sum_{i_2,\ldots,i_n\neq i} J_{ii_2\ldots i_n}\xi^1_i \xi^1_{i_2}\cdots \xi^1_{i_n} s_{i_1}\cdots s_{i_n})$, we note that $\Expv(m)=\Expv(m_i)$, so to find $\Expv(m)$ we just need to find the probability of one of the spins being retrieved correctly. Also, if we were instead estimating the capacity of the network using the techniques of equilibrium statistical physics (as done by Amit et al.~\cite{amit_storing_1985}), we would be quantifying retrieval using an order parameter which is analogous to $\Expv(m)$.
\begin{figure}
    \centering
    \includegraphics[width=0.5\linewidth]{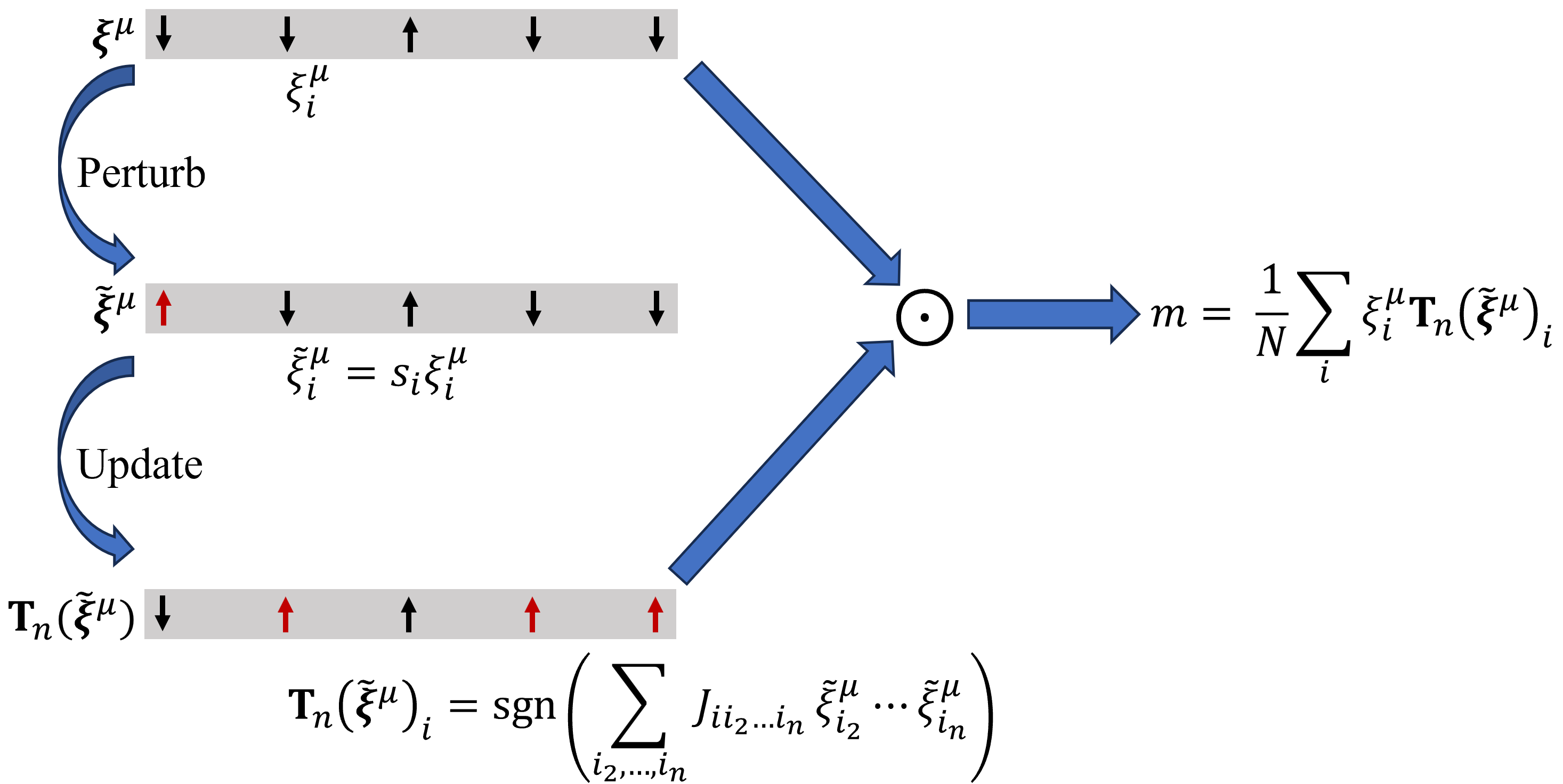}
    \caption{Schematic description of the procedure used to describe the retrieval accuracy for a modern Hopfield network with $N=5$ spins. We consider a pattern, say $\vb* \xi^\mu$ (where $\mu\in\qty{1,\ldots,K}$), and perturb it by randomly and independently flipping each spin $\xi^\mu_i$ with probability $\delta$. That is, the perturbed state is $\tilde{\vb* \xi}^\mu$, with components $\tilde \xi^\mu_i = s_i \xi^\mu_i$, where $s_i$ can independently take values $1$ and $-1$ with probabilities $1-\delta$ and $\delta$ respectively. In this figure, $\delta=0.2$. One step of the Hopfield update $\vb T_n$ is then applied on the perturbed pattern, where $n$ is the order of the interaction ($n=2$ corresponds to the usual Hopfield network). We find the overlap $m$ between the updated state $\vb T_n (\tilde{\vb* \xi^\mu})$ and the original pattern $\vb* \xi^\mu$, which is a measure of how accurately the pattern is retrieved after one step of the Hopfield update. $m$ is a random variable, so we can study its expected value or probability distribution. In this figure, we have $m=-0.2$ as only $2$ of the $5$ components of $\vb T_n(\tilde{\vb* \xi}^\mu)$ match with $\vb* \xi^\mu$. This is because the update protocol was based on the $4$-spin interaction model with synaptic noise described in Fig.~\ref{fig:quarticscheme}, for which the retrieval accuracy is strongly affected by the high perturbation rate $\delta = 0.2$ in this example.}
    \label{fig:iterate}
\end{figure}

An alternate measure for estimating retrieval would be to use probabilities instead of expected values. The probability $\Proby(m_i=1)$ of any particular spin being retrieved correctly is linearly related to $\Expv(m)$. We can also consider the probability $\Proby(m\ge m_0)$, of the return overlap being greater than some threshold value $m_0$. In particular, $\Proby(m=1)=\Proby(m_i=1\forall i)$ is the probability of perfect retrieval. This is a useful measure as in some contexts, we may care more whether the pattern is retrieved perfectly (or more accurately than some fixed threshold) or not, and in such cases it is useful to know the probability of ``successful'' retrieval under these criteria. This is, however, more difficult to estimate, as the states of the different spins after the Hopfield update are not independent, so the probability of perfect retrieval can not be simply written as $\Proby(m=1)=\prod_i \Proby(m_i=1)$. 
There is also no order parameter analogous to this probability when analyzing the behavior of the network using mean field theory.

We expect both $\Expv(m)$ and the probability of perfect retrieval $\Proby(m=1)$ to be close to $1$ for small number of patterns $K$ and decrease for larger $K$. We can then define the capacity to be the maximum $K$ for which the chosen measure of retrieval error remains below a particular tolerance threshold. It is common to choose the threshold to be $1/N$ (so it vanishes in the limit $N\to \infty$), so a commonly used definition of capacity is the number of patterns $K$ for which the probability of {\em im}perfect retrieval is $\Proby(\vb T\tilde{\vb*\xi}^1\neq\vb*\xi^1)=1/N$.

\section[{Extension to networks with n-spin interactions}]{Estimation of the retrieval accuracy and capacity}\label{dense}

Following the prescription of the previous section, here we compute the retrieval accuracy of the system after a single step of Hopfield update in a system with $n$-spin interaction for different noise models. In appendix~\ref{details} we provide further algebraic details behind the analysis, and in appendix~\ref{descriptionforbin} we work through the derivation of the results for the usual Hopfield network with two-spin couplings (in which case we obtain the same results as previous work, e.g., ref.~\cite{mceliece_capacity_1987})

It is convenient to define the random variable $X_i = \xi^1_i \sum_{i_2\ldots i_n} J_{ii_2\ldots i_n} \tilde \xi^1_{i_2}\ldots \tilde\xi^1_{i_n}$, which is positive if the $i$th spin is retrieved correctly and negative otherwise. Its mean and variance can be found to be
\begin{align}
    \Expv(X_i) &= \binom{N-1}{n-1} \mu_P (1-2\delta)^{n-1}, \\
    \Varnc(X_i) &= \binom{N-1}{n-1} K \mu_{P^2} + \binom{N-1}{n-1} \mu_{\eta^2} \mu_{P^2} - \binom{N-1}{n-1} \mu_P^2 (1-2\delta)^{2(n-1)} \nn\\
    &\quad + \mu_{P}^2 (1-2\delta)^{2(n-1)} \binom{N-1}{n-1} \sum_{c=1}^{n-2} \binom{N-n}{n-1-c} \binom{n-1}{c} \qty( (1-2\delta)^{-2c} - 1 ).
\end{align}
(These have been derived in detail in appendix~\ref{details}.) To find the distribution of the  $\sgn(X_i)$, we assume that the distribution of $X_i$ is  Gaussian (comparison with numerical simulations reveals this to be a good approximation, even though the simplest form of the central limit theorem is not applicable here as $X_i$ is the sum of random variables which are not all independent and identically distributed). Then we find that
\begin{align}
    \Proby(X_i > 0) &= \frac{1}{2} + \frac{1}{2} \erf(\frac{\Expv(X_i)}{\sqrt{2\Varnc(X_i)}}) \nn\\
    &= \frac{1}{2} + \frac{1}{2} \erf( \sqrt{\frac{\binom{N-1}{n-1}}{2\qty[ (K+\mu_{\eta^2})(1-2\delta)^{-2(n-1)}\mu_{P^2}/\mu_P^2 - 1 + \sum_{c=1}^{n-2} \binom{N-n}{n-1-c}\binom{n-1}{c} \qty( (1-2\delta)^{-2c} - 1 )]}} ),
\end{align}
and the overlap with the pattern has expected value
\begin{align}
    \Expv(m) &= \Expv(\sgn(X_i)) = \erf( \sqrt{\frac{\binom{N-1}{n-1}}{2\qty[ (K+\mu_{\eta^2})(1-2\delta)^{-2(n-1)}\mu_{P^2}/\mu_P^2 - 1 + \sum_{c=1}^{n-2} \binom{N-n}{n-1-c}\binom{n-1}{c} \qty( (1-2\delta)^{-2c} - 1 )]}} ).\label{lessapprox}
\end{align}
Since we are considering the case where $N \gg 1$, and if we assume $K\sim N^{n-1}$, this can be approximated as
\begin{align}
    \Expv(m) &= \erf( (1-2\delta)^{n-1}\sqrt{\frac{\binom{N-1}{n-1}\mu_P^2}{2\mu_{P^2}(K+\mu_{\eta^2})}} ).\label{yesnoise}
\end{align}

\subsection{Noise-free Hebbian interactions}
Let us first consider the special case where the interactions are Hebbian with no noise (i.e., $\eta_{i_1\ldots i_n} = 0$ and $P_{i_1\ldots i_n} = 1$ with probability $1$), in which case we have
\begin{align}
    \Expv(m) &= \erf( (1-2\delta)^{n-1}\sqrt{\frac{\binom{N-1}{n-1}}{2K}} ) = \erf( \frac{(1-2\delta)^{n-1}}{\sqrt{2(n-1)!\alpha_n}} ),\label{nonoise}
\end{align}
where $\alpha_n = K/[(N-1)\cdots(N-n+1)]\approx K/N^{n-1}$ (if $N\gg n^2$). Therefore, if we define the capacity of the model as the number of patterns for which the expected overlap is above some threshold value of $m_0$, then the capacity $K_c = [(1-2\delta)^{2(n-1)}/2\erf^{-1}(m_0)^2]\binom{N-1}{n-1}$ will scale with the system size as $N^{n-1}$.

We know that the error function for $x \ll 1$ is approximately linear, $\erf(x) \approx 2x/\sqrt{\pi}$ , and is close to $1$ and almost flat  for $x \gg 1$, $\erf(x) \approx 1-e^{-x^2}/x\sqrt{\pi}$. Therefore, the expected overlap $\Expv(m)$ approaches 1 when $K=1$ (assuming $\delta$ is small enough),  and decays as $K^{-1/2}$ when $K$ is of the order of $N^{n-1}$ or larger. The factor of $(1-2\delta)^{n-1}$ in the argument of the error function implies that that the retrieval accuracy depends strongly on deviation $\delta$ of the perturbed initial state from the perfect pattern, in particular for higger interaction orders $n$. 
For $K< [(1-2\delta)N]^{n-1}$ the expected overlap has plateau at $1$, before falling as $K$ is increased. We verify this numerically in section~\ref{numerics}. 

\subsection{Noisy interactions}

Including the additive and multiplicative noise terms $\eta_{i_1\ldots i_n}$ and $P_{i_1\ldots i_n}$ affects the expected overlap $\Expv(m)$ by changing the denominator  of the error function argument. The additive noise effectively replaces the number of patterns $K$ in in eq.~\eqref{yesnoise} with $K+\mu_{\eta^2}$, where $\mu_{\eta^2}$ is nonnegative, and zero if and only if there is zero additive noise. This affects the retrieval accuracy the same way as adding more patterns would, and significantly affects the capacity only if $\mu_{\eta^2}$ is of the order of $N^{n-1}$ or larger. Thus not only can the modern Hopfield networks store many more patterns than the original $n=2$ model, but they are also more robust to additive noise in their interactions. We also note that, at least in the cases where the additive noise is constant or Gaussian, the retrieval accuracy only depends on its second order raw moment $\mu_{\eta^2}$. Therefore, a constant shift, a Gaussian noise centered at $0$, and a Gaussian noise with an offset will affect the retrieval accuracy identically as long as they have the same $\mu_{\eta^2}$.

The multiplicative noise, on the other hand, multiplies the denominator in the argument of the error function in eq.~\eqref{yesnoise} by a factor of $\mu_{P^2}/\mu_P^2$, which is $1$ if $P_{i_1\ldots i_n}$ is a degenerate random variable (i.e., multiplying the interaction coefficients by a constant does not affect the retrieval accuracy) and greater than $1$ otherwise. This results in the capacity being reduced by a factor of $\mu_{P^2}/\mu_P^2$. As a special case, we can consider the situation where $P_{i_1\ldots i_n}$ are i.i.d.\ Bernoulli random variables with probability $p$ of being $1$ and $1-p$ being 0; this corresponds to the scenario where some interaction terms have been randomly and independently deleted with probability $1-p$. In this case, we see that $\mu_P=\mu_{P^2}=p$, so the denominator of the error function gains a factor of $1/p$ and the capacity of the system is scaled by a factor of $p$. That is, the capacity is proportional to the fraction of the retained coupling constants, which could be expected qualitatively.

\subsection{Clipped couplings}
In this section study the model where the interaction constants have all been constrained to be $\pm 1$ using a sign function, 
\begin{align}
\check{J}_{i_1\ldots i_n}= \sgn\left[ \qty(\sum_\mu \xi^\mu_{i_1}\cdots \xi^\mu_{i_n} + \eta_{i_1\ldots i_n})P_{i_1\ldots i_n} \right].
\end{align}

The projection to binary values does not itself introduce extra noise. However, the way noises affect the capacity is different than in the Hebbian case.
We take the sign of the interaction term \emph{after} introducing the additive and multiplicative noise to ensure that the interaction coefficients retain the magnitude $1$ (or $0$) even with noise. Therefore, the additive noise $\eta_{i_1\ldots i_n}$ in this case effectively flips the sign of some of the interaction coefficients, with weaker interactions being more likely to be flipped. Only the sign of the multiplicative noise factor $P_{i_1\ldots i_n}$ matters, so we will consider the cases where the only possible values of $P_{i_1\ldots i_n}$ are $0$ or $\pm 1$. In particular, if $P_{i_1\ldots i_n}\sim \text{Bernoulli}(p)$ are i.i.d.\ Bernoulli random variables, it corresponds to interactions being randomly deleted with probability $1-p$ as before. We can also consider the case where $P_{i_1\ldots i_n}$ takes the values $1$ and $-1$ with probability $q$ and $1-q$ respectively, which would be interpreted as the couplings being passed through a noisy channel which randomly flips the signs of some of the interactions.

We can similarly as before perturb a pattern, apply one step of the Hopfield update, and find its overlap with the original pattern. The variable determining whether the $i$-th spin is retrieved correctly in this case is
\begin{align}
\check X_i &= \sum_{\substack{1\le i_2 < \cdots< i_n\le N,\\ i_a \neq i \forall a}} \sgn( \sum_{\mu=1}^K \xi^1_i \xi^1_{i_2}\cdots \xi^1_{i_n} \xi^\mu_i \xi^\mu_{i_2}\cdots \xi^\mu_{i_n} P_{ii_2\ldots i_n}s_{i_2}\cdots s_{i_n} + \xi^1_i \xi^1_{i_1}\cdots \xi^1_{i_n} \eta_{ii_2\ldots i_n} P_{ii_2\ldots i_n}s_{i_2}\cdots s_{i_n} ) \nn\\
&= \sum_{\substack{1\le i_2 < \cdots< i_n\le N,\\ i_a \neq i \forall a}} \sgn( \sum_{\mu=1}^K \xi^1_i \xi^1_{i_2}\cdots \xi^1_{i_n} \xi^\mu_i \xi^\mu_{i_2}\cdots \xi^\mu_{i_n} P_{ii_2\ldots i_n}s_{i_2}\cdots s_{i_n} + \xi^1_i \xi^1_{i_1}\cdots \xi^1_{i_n} \eta_{ii_2\ldots i_n} P_{ii_2\ldots i_n}s_{i_2}\cdots s_{i_n} ) \nn\\
&\equiv \sum_{\qty{i_a}} \sgn(X_{ii_1\ldots i_n}).
\end{align}

The expectations and variances of the random variables appearing in the above expression can be found in the same manner as in the case where the interactions were not clipped, which used to find that
\begin{align}
\Expv(X_{ii_1\ldots i_n}) &= \mu_P (1-2\delta)^{n-1}, \\
\Varnc(X_{ii_1\ldots i_n}) &= (K+\mu_{\eta^2}) \mu_{P^2} - \mu_P^2 (1-2\delta)^{2(n-1)},
\end{align}
which implies that the expected value of the sign of this variable is
\begin{align}
\Expv(\sgn(X_{ii_1\ldots i_n})) = \erf( \frac{1}{\sqrt{ 2\mu_{P^2}(K+\mu_{\eta^2})/\mu_P^2(1-2\delta)^{2(n-1)} - 2 }} ) \equiv \check e.
\end{align}
Ignoring the correlations between $X_{ii_1\ldots i_n}$ for different sets $\qty{i_a}$, we can estimate that
\begin{align}
\Expv(\check{X}_i) &= \binom{N-1}{n-1} \check e, \\
\Varnc(\check X_i) &= \binom{N-1}{n-1} (1-\check e^2).
\end{align}
Therefore, the expected overlap is
\begin{align}
\Expv(\check m) &= \Expv(\sgn(\check X_i)) = \erf(\frac{\Expv(\check{X}_i)}{\sqrt{2\Varnc(\check X_i)}}) \nn\\
&= \erf( \sqrt{\frac{\binom{N-1}{n-1}\check e^2}{2(1-\check e^2)}} ).
\label{noisesign}
\end{align}
For $K\gg 1$, we have $\check e \approx \sqrt{ 2\mu_P^2(1-2\delta)^{2(n-1)}/\pi\mu_{P^2}(K+\mu_{\eta^2}) }\ll 1$, which implies
\begin{align}
    \Expv(\check m) &\approx \erf( \sqrt{\frac{(1-2\delta)^{2(n-1)}\mu_P^2\binom{N-1}{n-1}}{\pi \mu_{P^2}(K+\mu_{\eta^2})}} ).
\end{align}
Comparing this with the expression in eq.~\eqref{yesnoise} where the interaction coefficients were not clipped, we note that they differ by a factor of $2/\pi$ in the argument of the error function. Therefore, the capacity of this model is less than that of the unclipped model by a factor of $2/\pi$, but otherwise scales with system size the same way.

Therefore, we find that the capacity in this case will scale as $N^{n-1}$ (as found before in~\cite{krotov_dense_2016,agliari_tolerance_2020}). The capacity also depends strongly on the size of the required basin of attraction around each pattern; if we require patterns with a Hamming distance of up to $N\delta$ from a pattern to be retrieved, the capacity has a factor of $(1-2\delta)^{2(n-1)}$. Finally, if we use binary couplings, this introduces the same factor of $2/\pi$ in the capacity as had been previously obtained for the network with two-spin interactions~\cite{mceliece_capacity_1987}. Therefore, if we wish to design a network with a given capacity, the number of spins the network needs to have will only go up by a factor of $(\pi/2)^{1/(n-1)}$, which approaches $1$ for large $n$.

\section{Numerical simulations}\label{numerics}
We numerically simulated the dynamics of modern Hopfield networks using the NumPy library in Python3~\cite{harris_array_2020} to compare against their predicted retrieval accuracy. The memory patterns $\xi^\mu$ were generated as random vectors of  size $N$ (with every element equal to $\pm 1$), and the interaction coefficients $J_{i_1\ldots i_n}$ were computed using the Hebb rule and stored as an $n$ dimensional array. To determine a retrieval accuracy, we chose one of the  patterns, flipped the sign of each of its components independently with probability $\delta$, implemented one step of the Hopfield update on it using the stored interaction coefficients, and then found the overlap $m$ of the resulting vector with the original pattern. For every set of stored patterns, we repeated the above procedure 500--2000 times (using different starting patterns) to obtain a sample large enough to estimate statistical parameters while computing in a reasonable amount of time. We used the simulation data to estimate the average overlap $\Expv(m)$, and then repeated this whole procedure for varying numbers of patterns (keeping the system size $N$ and perturbation probability $\delta$ fixed) to obtain a plot of the average overlap as a function of the number of patterns using the Python library Matplotlib~\cite{hunter_matplotlib_2007}.

\begin{figure}
    \centering
    \includegraphics[width=0.9\linewidth]{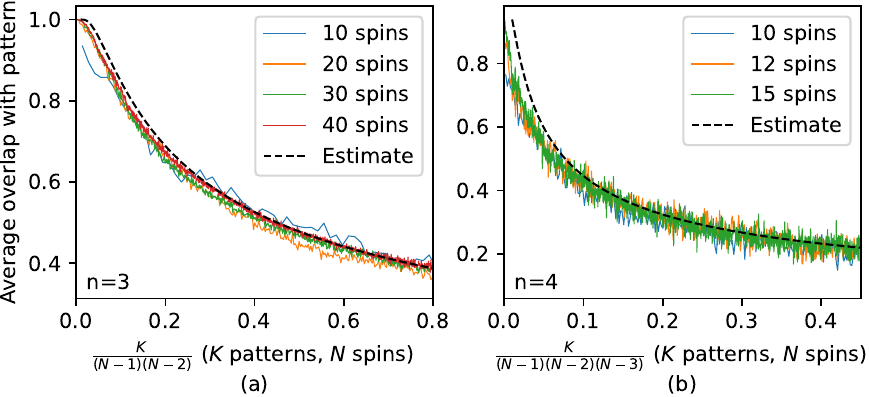}

    \caption{Pure Hebbian couplings. Average overlap of a pattern with itself  after randomly ``corrupting" $\delta = 0.1$ of the spins and doing one step of Hopfield update (in a modern Hopfield network with Hebbian $n$-body interactions) as a function of the scaled number of patterns $\alpha_n = K/(N-1)\cdots(N-n+1)$. The first plot (a) is for $n=3$ for which the simulations were run for $N=10$, $20$, $30$, and $40$ spins, and the second plot (b) is for $n=4$, for which the simulations were run for system sizes of $N=10$, $12$, and $15$. The analytical estimate obtained in eq.~\eqref{nonoise} has also been plotted as a black dashed line. For $\delta\neq 0$, the expression in eq.~\eqref{lessapprox} is a better approximation for large system size $N$ and number of patterns $K$ but deviates somewhat from the simulation results for smaller $K$, as can be seen in both plots and is more apparent in plot (b).}
    \label{fig:nonoise}
\end{figure}

The analytical expression obtained in eq.~\eqref{nonoise} was also calculated using the Python library Scipy~\cite{virtanen_scipy_2020} and plotted  for comparison with the simulation results. Because this expression depends on $K$ and $N$ only via $\alpha_n = K/(N-1)\cdots(N-n+1)$, we expect plots of $\Expv(m)$ vs.\ $\alpha_n$ to agree with this expression for all sufficiently large values of $N$ (with the possible exception when $\alpha_n \ll 1$, in which case the approximation used to obtain eqs.~\eqref{yesnoise} and \eqref{nonoise} (which assumed $K \sim N^{n-1}$) is not valid). 
In figure~\ref{fig:nonoise}, we plot the average overlap $\Expv(m)$ vs.\ the scaled number of patterns $\alpha_n = K/(N-1)\cdots(N-n+1)$ (keeping $N$ fixed and varying $K$ for each plot) for two sets of modern Hopfield networks. The first plot shows the results for $N=10, 20, 30$ spins forming a modern Hopfield network with $n=3$-spin interactions,, whereas the second plot is for $N=10,12,15$ spins interacting vis $n=4$-spin interactions. In each case, the perturbation rate is $\delta = 0.1$. We find results for different $N$ to be in agreement with each other and  with the  estimate given by eq.~\eqref{nonoise}, except for smaller values $\alpha_3 < 0.3$ and $\alpha_4<0.1$. In the first plot, we note that the average overlap plateaus near $1$ for very small $\alpha_3$ and then falls off with $\alpha_3$ as expected. For the second plot, because of the smaller system sizes, the overlap is less than $1$ even for very few memories and falls off with $K$ without a plateau.

\begin{figure}
    \centering
    \includegraphics[width=0.45\linewidth]{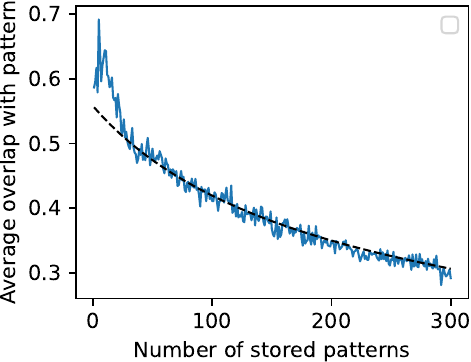}
    \caption{Noisy Hebbian couplings. Average overlap of a pattern with itself as a function of the number of stored patterns after randomly ``corrupting'' $\delta=0.1$ of the spins doing one step of Hopfield update (in a modern Hopfield network with $N=20$ spins interacting via $n=3$-body interactions). The interaction coefficients are defined via the Hebb rule, but each coefficient also has an independent additive noise factor following the Gaussian distribution $N(\mu=0,\sigma^2=100)$, and an independent multiplicative factor following the Bernoulli distribution $B(p=0.9)$. The latter corresponds to $10\%$ of the interaction terms being randomly deleted. The analytical estimate obtained in eq.~\eqref{yesnoise} has also been plotted as a black dashed line.}
    \label{fig:yesnoise}
\end{figure}

Similarly, we simulated networks where the interaction coefficients had additive and/or multiplicative noise. In particular, we simulated the cases where the additive term on each interaction coefficient was a normal random variable centered at $0$, and the multiplicative factor was a Bernoulli random variable taking the value $1$ with probability $p$ and $0$ otherwise. Figure~\ref{fig:yesnoise} plots the expected overlap vs the scaled number of patterns $\alpha_3$ for a modern Hopfield network of size $N=20$ with $n=3$-spin interactions, where the additive noise is Gaussian with mean $0$ and variance $100$, and the multiplicative noise follows a Bernoulli distribution with parameter $p=0.9$ (interpreted as retaining $90\%$ the interaction terms and deleting the rest). The plots agree with the estimate given by eq.~\ref{yesnoise}, also plotted in black. Because of the high noise, the expected overlap is low (around $0.6$) even for very few stored patterns and falls off with the number of memories with no plateau. 

\begin{figure}
    \centering
    
    \includegraphics[width=0.9\linewidth]{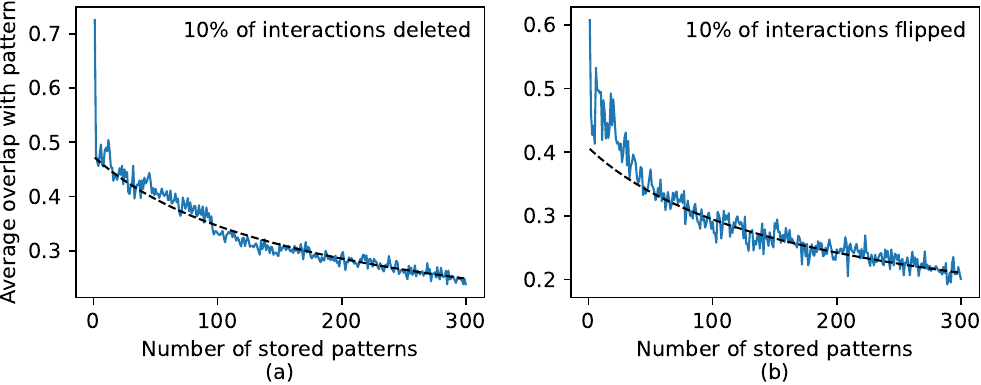}
    \caption{Clipped noisy Hebbian couplings. Average overlap of a pattern with itself as a function of the number of stored patterns after randomly ``corrupting'' $\delta=0.1$ of the spins doing one step of Hopfield update (in a modern Hopfield network with $N=20$ spins interacting via $n=3$-body interactions). The interaction coefficients are initially defined via the Hebb rule, but each coefficient is also perturbed by an independent additive noise factor following the Gaussian distribution $N(\mu=0,\sigma^2=100)$, and then clipped to have magnitude $1$. The two plots correspond to systems where the interaction coefficients are (a) randomly deleted with probability $0.1$, or (b) randomly flipped in sign with probability $0.1$. The analytical estimate obtained in eq.~\eqref{noisesign} has also been plotted as a black dashed line in each case.}
    \label{fig:noisesign}
\end{figure}

Networks with clipped interactions (both in presence and absence of noise) were simulated by defining a new array of the clipped couplings $\check J_{i_1\ldots i_n} = \sgn(J_{i_1\ldots i_n})$. Figure~\ref{fig:noisesign} shows the plots of the expected overlap vs.\ $\alpha_3$ in modern Hopfield networks of system size $N=20$ with $3$-spin interactions, where the initial perturbation is $\delta=0.1$. Each interaction coefficient was perturbed with an additive Gaussian noise with mean $0$ and variance $100$ and then clipped to have magnitude $1$. The first plot corresponds to the case where $10\%$ of the interactions were randomly deleted, and the second is when $10\%$ of the interactions were randomly flipped in sign. The plots are found to agree with the estimates given by eq.~\eqref{noisesign}. 

\color{black}\section{Discussions, applications and outlook}\label{outlook}

A model of associative memory capable of retrieving $K$ patterns $\vb*\xi^\mu$, each containing $N$ binary neurons $\xi^\mu_i$, stores $NK$ bits of information. Equation~\eqref{nham} expresses the Hamiltonian of the modern Hopfield network in terms of the $NK$ pattern components $\xi^\mu_i$, in which case the pattern components themselves are the parameters of the model (and we need to introduce new parameters if more patterns need to be stored). While other ways to formulate and parametrize such models have been recently proposed (such as using random features~\cite{hoover_dense_2024}), a possible way to implement this dynamics in a physical or biological system is via interactions between the constituent neurons, similar to the interactions between spins in the Ising model. Such interactions $J_{i_1\ldots i_n}$ involve two spins each ($n = 2$) for the standard Hopfield network and become many-body ($n> 2$) for the dense associative memory models considered here; typically they are constructed from the stored patterns using the Hebb rule. If we exclude self-interactions, the model is parametrized by $\binom{N}{n}$ interaction coefficients, each of which can take $K+1$ possible values. It is of interest to investigate how resilient is the capacity of the network is to noise in these interaction coefficients, as well as to ``coarsening'' of the resolution in their values, as this determines the ability of such networks to successfully retrieve patterns in systems such as the brain.

We found that the effect of additive noise on the capacity is insignificant as long as the noise $\eta_{i_1\ldots i_n}$ scales with the system size $N$ at a power of less than $(n-1)/2$; specifically, if the mean squared noise satisfies $\Expv(\eta_{i_1\ldots i_n}^2)\ll N^{n-1}$, then the capacity is unaffected. This constitutes the level of robustness of the network to extrinsic noise. Additive noise scaling as $\Expv(\eta_{i_1\ldots i_n}^2)\sim N^{n-1}$ reduces the capacity by a constant factor without changing its dependence on system size, but for larger noise (such as when $\Expv(\eta_{i_1\ldots i_n}^2)\sim N^b$ with $b>n-1$) the system fails to retrieve in the limit of large system size irrespective of the number of patterns.

Another possibility to consider is when the additive noise scales with the number of patterns being stored. This may happen, for example, if,  when training the interaction coefficients via he Hebb rule, an independent noise term gets introduced for every pattern being stored. In this case $\eta_{i_1\ldots i_n} = \sum_\mu \eta^\mu_{i_1\ldots i_n}$, where each of these terms can be thought to be i.i.d.\ with mean squared value $\tilde \mu_{\eta^2}$, which Agliari~\cite{agliari_tolerance_2020} calls noisy storing. In this case $\mu_{\eta^2} = K\tilde \mu_{\eta^2}$, and eq.~\eqref{yesnoise} has the term $K+\mu_{\eta^2} = K(1+\tilde\mu_{\eta^2})$. For $\tilde\mu_{\eta^2}<<1$ the effect of the additive noise is once again negligible, and for finite $\tilde\mu_{\eta^2}$ that does not depend on the system size the capacity is reduced by a factor of $1+\tilde\mu_{\eta^2}$ but otherwise scales with system size as $N^{n-1}$ as usual. If the additive noise per pattern diverges with system size as $\tilde\mu_{\eta^2}\sim N_b$, then the scaling behavior of the capacity changes to $K_\text{max} \sim N^{n-1-b}$, and thus the system can retrieve patterns in the thermodynamic limit only if $b\le n-1$.

Multiplicative noise, on the other hand, is found to reduce the capacity without changing its scaling behavior with system size, irrespective of the magnitude of the noise (as long as its expectation is positive). In particular, if only a random subset of interactions is retained and the rest are set to zero (so each interaction has a probability $p$ of being retained), then the capacity is reduced by a factor of $p$. Such a diluted network can thus still function as a dense associative memory, even if with a smaller capacity. A network where each interaction is retained with probability $q$ and otherwise flipped in sign can still retrieve some patterns as long as $q>1/2$, and in this case the capacity is reduced by a factor of $(2q-1)^2$.

We also considered networks where the interaction strengths had been clipped to all have the same magnitude and only the signs differ. We find that even with such drastic reduction in the precision of the interaction coefficients, the capacity is reduced by only a factor of $2/\pi$. This indicates that a physical system can be engineered to behave as a dense associative memory even if the interaction strengths are rounded to have lower ``resolution" than the Hebbian values. If we simulate such a system digitally, each interaction coefficients can be stored in a single bit, more compactly than the Hebbian values which require $\log_2(K+1)$ bits. 
If we wish to retrieve a fixed number $K$ of patterns using the smallest possible modern Hopfield network with $n$-spin interactions, and simulate the interactions digitally, then it turns out that for sufficiently large $K$, using clipped interactions would require storing fewer bits overall for the interaction coefficients, even taking into account the slightly larger number of spins required in the clipped network due to the reduced capacity. This is because the number of spins in the clipped model is larger by a factor of $(\pi/2)^{1/(n-1)}$, so the ratio of the required number of bits in the unclipped and clipped networks is $\log_2(K+1)/(\pi/2)^{n/(n-1)}$, which is greater than $1$ for all interaction orders $n$ as long as $K\ge 5$ (which always holds as we are assuming $N\gg 1$ and $K\sim N^{n-1}$). We did not consider cases where the interaction strengths are clipped less strongly and are allowed to have more than two different values, but we expect such networks to have capacities even closer to the unclipped Hebbian network.

We thus conclude that modern Hopfield networks defined via $n$-spin interactions can retrieve patterns in many cases even when the interactions have additive or multiplicative noise or are clipped to have lower precision compared to the Hebbian values. The capacity of the networks grows as the $(n-1)$-th power the system size in all cases, though they may be reduced by a factor depending on the noise or the level of clipping. The only cases where we found that the network may fail to retrieve are in the presence of additive noise growing with the system size faster than its $(n-1)$-th power, and in the presence of multiplicative noise whose expectation is nonpositive.

While we considered identically distributed additive and multiplicative noise terms affecting every interaction coefficient independently, further work can be done to explore the effects of other kinds of noise, perhaps physically motivated by the process by which the network learns and stores the interactions. For example, Agliari et al. have considered the effects of other forms of additive noise, which they associate with noisy patterns, learning, or storing~\cite{agliari_tolerance_2020}. We focused on the retrieval accuracy after one step of the synchronous Hopfield update, but it may be interesting to study the behavior of the network after multiple steps of stochastic dynamics given by a nonzero pseudotemperature, which may be analyzed using tools from equilibrium statistical physics such as replica theory, such as the analysis for the networks with pure Hebbian interactions carried out in~\cite{albanese_replica_2022}, for very noisy networks with linear storage capacity in~\cite{agliari_neural_2020, alemanno_interpolating_2020} and for general classical and quantum neural networks in~\cite{barney_neural_2024}. The effect of nonequilibrium dynamics on the capacity, as discussed for classic Hopfield networks in~\cite{behera_enhanced_2023, behera_correspondence_2024}, is also worth studying. It may also be of interest to consider networks with correlated patterns such as those considered in~\cite{marzo_effect_2022, burns_semantically-correlated_2024}, as well as systems with continuous components, such as the models discussed in~\cite{ramsauer_hopfield_2021, karin_enhancernet_2024, grishechkin_hierarchical_2025}, instead of the Ising systems studied here.

\color{black}
\section*{Acknowledgments}
We thank Ilya Nemenman for useful discussions. This work was supported by the US Department of Energy, Office of Science, Basic Energy Sciences, Materials Sciences and Engineering Division.

\printbibliography

\appendix

\section{Details for the derivation of the expected overlap}\label{details}
Here we describe in more detail the derivation of the expected overlap $\Expv(m)$ for the modern Hopfield network. We consider a network with $N$ spins interacting via $n$-body interactions, and storing $K$ patterns $\vb*\xi^\mu$ whose components $\xi^\mu_i$ are i.i.d.\ Rademacher variables. Let us assume that we start with the $\mu$-th pattern $\vb*\xi^\mu$ and perturb it by flipping each of its components $\xi^\mu_i$ with probability $\delta$. The resulting state $\tilde{\vb*{\xi}}^\mu$ has components $\tilde \xi^\mu_i = \xi^\mu_i s_i$, where $s_i$ are i.i.d.\ discrete random variables taking the values $1$ and $-1$ with probabilities $1-\delta$ and $\delta$ respectively. (Therefore we have $\Expv(s_i)=1-2\delta$, and $s_i^2=1$ with probability $1$.) We perform one step of the Hopfield update defined in eq.~\eqref{update} to obtain the state $\vb T_n \tilde{\vb* \xi}^\mu$ with components
\begin{align}
    (\vb T_n \tilde{\vb* \xi}^\mu)_i &= \sgn(\sum_{\substack{1\le i_2\le\cdots i_n\le N \\ i_\alpha \neq i \forall \alpha}} J_{ii_2\ldots i_n} \tilde \xi^\mu_{i_2} \ldots \tilde \xi^\mu_{i_n}) = \sgn(\sum_{\substack{1\le i_2\le\cdots i_n\le N \\ i_\alpha \neq i \forall \alpha}} J_{ii_2\ldots i_n} \xi^\mu_{i_2} \ldots \xi^\mu_{i_n} s_{i_2}\cdots s_{i_n}),
\end{align}
where $J_{j_1\ldots j_n}$ are the $n$-body interaction coefficients. We assume that they follow the Hebb rule but may have additive or multiplicative noise, and thus are defined as
\begin{align}
    J_{j_1\ldots j_n} = P_{j_1\ldots j_n} \qty(\sum_\nu \xi^\nu_{j_1}\cdots \xi^\nu_{j_n} + \eta_{j_1\ldots j_n}),
\end{align}
where the additive noise terms $\eta_{j_1\ldots j_n}$ are independent and identically distributed, and so are the multiplicative noise factors $P_{j_1\ldots j_n}$. We can then consider the variable
\begin{align}
    X_i & = \sum_{\substack{1\le i_2\le\cdots i_n\le N \\ i_\alpha \neq i \forall \alpha}} P_{ii_2\ldots i_n} \qty(\sum_\nu \xi^\nu_i\xi^\nu_{i_2}\cdots \xi^\nu_{i_n} + \eta_{ii_2\ldots i_n}) \xi^\mu_i \xi^\mu_{i_2} \ldots \xi^\mu_{i_n} s_{i_2}\cdots s_{i_n} \nn\\
    &= \sum_{\nu, \qty{i_\alpha}} Y^\nu_{i\qty{i_\alpha}} + \sum_{\qty{i_\alpha}} Z_{i\qty{i_\alpha}},
\end{align}
where
\begin{align}
    Y^\nu_{i\qty{i_\alpha}} &= \xi^\mu_i \xi^\mu_{i_2} \ldots \xi^\mu_{i_n} \xi^\nu_i\xi^\nu_{i_2}\cdots \xi^\nu_{i_n} \xi^\mu_i \xi^\mu_{i_2} \ldots \xi^\mu_{i_n} P_{ii_2\ldots i_n} s_{i_2}\cdots s_{i_n}, \\
    Z_{i\qty{i_\alpha}} &= \xi^\mu_i \xi^\mu_{i_2} \ldots \xi^\mu_{i_n} P_{ii_2\ldots i_n} \eta_{ii_2\ldots i_n}) s_{i_2}\cdots s_{i_n},
\end{align}
and $\qty{i_\alpha}$ is shorthand for $i_2,\ldots, i_n$. The sign of $X_i$ will be $\sgn(X_i)=\xi^\mu_i (\vb T_n \vb* \xi^\mu)_i$, which determines whether the $i$-th spin is correctly retrieved, and the overlap of the state with the original pattern is $\langle \vb* \xi^\mu, \vb T_n \tilde{\vb*\xi}^\mu \rangle = \sum_i \sgn(X_i)/N$. Therefore, the statistical properties of $X_i$ can be used to estimate the retrieval accuracy.

Because $\xi^\mu_i$ are i.i.d.\ Rademacher variables, we have $\Expv(\xi^\mu_i)=0$, and ${\xi^\mu_i}^2=1$ with probability $1$. Therefore, any term involving a product of $\xi^\mu_i$ factors will have nonzero expectation only if they  ``pair up''. We then find that
\begin{align}
    \Expv(Y^\nu_{i\qty{i_\alpha}}) &= \delta_{\mu\nu} \mu_{P} (1-2\delta)^{n-1},\\
    \Expv(Z_{i\qty{i_\alpha}}) &= 0.
\end{align}
Since there are $\binom{N-1}{n-1}$ choices for the components $\qty{i_2,\ldots,i_n}\subset\qty{1,\ldots,N}\setminus\qty{i}$, we conclude that $\Expv(X_i) = \binom{N-1}{n-1}\mu_{P} (1-2\delta)^{n-1}$.

To find the variance of $X_i$, we need to find the variances as well as pairwise covariances of the terms $Y^\nu_{i\qty{i_\alpha}}$ and $Z_{i\qty{i_\alpha}}$. We note that 
\begin{align}
    \Covar(Z_{i\qty{i_\alpha}}, Z_{i\qty{j_\beta}}) &= \begin{cases}
        \Varnc(Z_{i\qty{i_\alpha}}) = \mu_{P^2}\mu_{\eta^2}, &\text{if } \qty{i_\alpha} = \qty{j_\beta}, \\
        0, &\text{otherwise},
    \end{cases} \\
    \Covar(Z_{i\qty{i_\alpha}}, Y^\nu_{i\qty{j_\beta}}) &= 0.
\end{align}
For finding the covariance $\Covar(Y^\nu_{i\qty{i_\alpha}}, Y^\lambda_{i\qty{j_\beta}})$, let us first consider the case where $\qty{i_\alpha}=\qty{j_\beta}$. In this case, the covariance $\Covar(Y^\nu_{i\qty{i_\alpha}}, Y^\lambda_{i\qty{i_\alpha}})$ is $\mu_{P^2}$ if $\nu=\lambda\neq \mu$, $\mu_{P^2}-\mu_{P}^2(1-2\delta)^{2(n-1)}$ if $\nu=\lambda=\mu$, and $0$ if $\nu\neq\lambda$. However, if $\qty{i_\alpha}\neq \qty{j_\beta}$, the covariance can be nonzero only if $\nu=\lambda=\mu$. In this case, we have $\Covar(Y^\mu_{i\qty{i_\alpha}}, Y^\mu_{i\qty{j_\beta}}) = \mu_{P}^2((1-2\delta)^{2(n-1-c)}-(1-2\delta)^{2(n-1)})$, where $0\le c=\abs{\qty{i_\alpha}\cap\qty{j_\beta}} \le n-2$ is the number of spins in common between the sets $\qty{i_\alpha}$ and $\qty{j_\beta}$. (We note that this covariance is zero when $c=0$, as well as for all $c$ in the case of no initial perturbation $\delta=0$.) The number of possible ways of sampling the sets $\qty{i_\alpha},\qty{j_\beta}\subset\qty{1,\ldots,N}\setminus\qty{i}$ with $\abs{\qty{i_\alpha}}=\abs{\qty{j_\beta}}=n-1$ and $\abs{\qty{i_\alpha}\cap\qty{j_\beta}}=c$ is $\binom{N-1}{n-1}\binom{n-1}{c}\binom{N-n}{n-1-c}$. Combining all the covariances, we find that
\begin{align}
    \Varnc(X_i) &= \binom{N-1}{n-1}\qty[\mu_{P^2} (K+\mu_{\eta^2})-\mu_P^2(1-2\delta)^{2(n-1)}] \nn\\
    &\quad + \binom{N-1}{n-1}\mu_P^2 (1-2\delta)^{2(n-1)} \sum_{c=1}^{n-2} \binom{n-1}{c}\binom{N-n}{n-1-c} \qty((1-2\delta)^{-2c}-1).
\end{align}
As we shall show below, the capacity of this network is of the order of $N^{n-1}$; when the number of memories is of that order, the first term in the above expression of the variance scales with system size as $N^{2(n-1)}$. The term on the second line, however, is of order $N^{2n-3}$ (corresponding to the term in the sum with $c=1$) and thus can be ignored when $N\gg 1$ and $K\sim N^{n-1}$, which is the regime we will be interested in. (This term also vanishes in the case of no initial perturbation, $\delta=0$.) To make further progress, we need to assume that the probability distribution of $X_i$ is approximately Gaussian. For any Gaussian random variable $G\sim N(\mu,\sigma^2)$, the probability of it being positive is $\Proby(G>0) = (1+\erf(\mu/\sqrt{2}\sigma))/2$. Therefore, the expected value of $\sgn(X_i)$ is estimated to be
\begin{align}
    \Expv(\sgn(X_i)) &= \erf(\frac{\Expv(X_i)}{\sqrt{2\Varnc(X_i)}}) \nn\\
    &\approx \erf(\frac{\binom{N-1}{n-1}\mu_{P} (1-2\delta)^{n-1}}{\sqrt{2\qty[\binom{N-1}{n-1}(\mu_{P^2} (K+\mu_{\eta^2})-\mu_P^2(1-2\delta)^{2(n-1)})]}})
\end{align}

While the cases where the interactions have additive and multiplicative noise can be treated together, we need to separately analyze the case where the interactions are also clipped. In this case we the interaction coefficients are defined as
\begin{align}
    \check J_{j_1\ldots j_n} = \sgn(P_{j_1\ldots j_n} \qty(\sum_\nu \xi^\nu_{j_1}\cdots \xi^\nu_{j_n} + \eta_{j_1\ldots j_n})),
\end{align}
i.e., similar to the previous case but with an additional sign function for the clipping. The analog of the variable $X_i$, determining whether the $i$-th spin is retrieved correctly, is now
\begin{align}
    \check X_i &= \sum_{\qty{i_\alpha}} \check J_{i\qty{i_\alpha}} \xi^\mu_i \qty(\prod_\alpha \xi^\mu_{i_\alpha} s_{i_\alpha}) \nn\\
    &= \sum_{\qty{i_\alpha}} \sgn(P_{i\qty{i_\alpha}} \sum_\nu \xi^\mu_i \xi^\nu_i \prod_\alpha\qty(\xi^\mu_{i_\alpha}\xi^\nu_{i_\alpha} s_{i_\alpha}) + P_{i\qty{i_\alpha}}\eta_{i\qty{i_\alpha}}\xi^\mu_i \prod_\alpha\qty(\xi^\mu_{i_\alpha}s_{i_\alpha})) \equiv \sum_{\qty{i_\alpha}} \sgn(X_{i\qty{i_\alpha}}),
\end{align}
where we now need to find the probability of each of the terms
\begin{align}
    X_{i\qty{i_\alpha}} &= P_{i\qty{i_\alpha}} \sum_\nu \xi^\mu_i \xi^\nu_i \prod_\alpha\qty(\xi^\mu_{i_\alpha}\xi^\nu_{i_\alpha} s_{i_\alpha}) + P_{i\qty{i_\alpha}}\eta_{i\qty{i_\alpha}}\xi^\mu_i \prod_\alpha\qty(\xi^\mu_{i_\alpha}s_{i_\alpha})
\end{align}
being positive. By similar arguments as above, we can find that
\begin{align}
    \Expv(X_{i\qty{i_\alpha}}) &= \mu_P (1-2\delta)^{n-1},
\end{align}
and
\begin{align}
    \Varnc(X_{i\qty{i_\alpha}}) &= (K+\mu_{\eta^2}) \mu_{P^2} - \mu_P^2 (1-2\delta)^{2(n-1)},
\end{align}
which implies
\begin{align}
    \Expv(X_{i\qty{i_\alpha}}) &= \erf(\frac{\mu_P (1-2\delta)^{n-1}}{\sqrt{2[(K+\mu_{\eta^2}) \mu_{P^2} - \mu_P^2 (1-2\delta)^{2(n-1)}]}}) \nn\\
    &\approx \frac{2\mu_P (1-2\delta)^{n-1}}{\sqrt{2\pi(K+\mu_{\eta^2}) \mu_{P^2}}}\quad(\text{for } K\gg 1),
\end{align}
and $\Varnc(\sgn(X_{i\qty{i_\alpha}})) = 1-E(\sgn(X_{i\qty{i_\alpha}}))^2$. Then $\check X_i$ satisfies
\begin{align}
    \Expv(\check X_i) &= \sum_{\qty{i_\alpha}} \Expv(\sgn(X_{i\qty{i_\alpha}})) = \binom{N-1}{n-1} \erf(\frac{\mu_P (1-2\delta)^{n-1}}{\sqrt{2(K+\mu_{\eta^2}) \mu_{P^2} - 2\mu_P^2 (1-2\delta)^{2(n-1)}}}) \nn\\
    &\approx \binom{N-1}{n-1} \frac{2\mu_P (1-2\delta)^{n-1}}{\sqrt{2\pi(K+\mu_{\eta^2}) \mu_{P^2}}},
\end{align}
and
\begin{align}
    \Varnc(\check X_i) &= \sum_{\qty{i_\alpha}} \Varnc(\sgn(X_{i\qty{i_\alpha}})) \nn\\
    &= \binom{N-1}{n-1}\qty(1 - \erf(\frac{\mu_P (1-2\delta)^{n-1}}{\sqrt{2(K+\mu_{\eta^2}) \mu_{P^2} - 2\mu_P^2 (1-2\delta)^{2(n-1)}}})^2) \nn\\
    &\approx \binom{N-1}{n-1} \qty[1 - \qty(\frac{2\mu_P (1-2\delta)^{n-1}}{\sqrt{2\pi(K+\mu_{\eta^2}) \mu_{P^2}}})^2],
\end{align}
where we have ignored the covariances $\Covar(\sgn(X_{i\qty{i_\alpha}}), \sgn(X_{i\qty{j_\beta}}))$.
Therefore, the expected overlap is
\begin{align}
    \Expv(\check m) &= \Expv(\sgn(\check X_i)) \approx \erf(\frac{\binom{N-1}{n-1} \frac{2\mu_P (1-2\delta)^{n-1}}{\sqrt{2\pi(K+\mu_{\eta^2}) \mu_{P^2}}}}{\sqrt{2\binom{N-1}{n-1} \qty[1 - \qty(\frac{2\mu_P (1-2\delta)^{n-1}}{\sqrt{2\pi(K+\mu_{\eta^2}) \mu_{P^2}}})^2]}}) \nn\\
    &= \erf((1-2\delta)^{n-1}\sqrt{\frac{\binom{N-1}{n-1}\mu_{P}^2}{\pi\mu_{P^2} (K+\mu_{\eta^2})}})
\end{align}

\section{Expressing dense associative memory Hamiltonians involving higher powers of overlaps in terms of multi-spin interactions and vice versa}\label{funfact}
Much of the work on dense associative memories, including that by Krotov~\cite{krotov_dense_2016}, define the Hamiltonians of such models in terms of the overlaps of the state with the patterns being stored. Therefore, if the system is in the state $\vb*\sigma$ (with components $\sigma_i$) and we wish to store the patterns $\vb*\xi^\mu$, then the Hamiltonian of the system can be expressed as
\begin{align}
H = -N \sum_\mu F(m_\mu), \label{krotovham}
\end{align}
where $m_\mu = \sum_i \xi^\mu_i \sigma_i /N$ are the overlaps. For the quadratic function $F(x)=x^2$, this corresponds to the standard Hopfield network, whereas $F(x)$ involving higher powers of $x$ indicates a dense associative memory.

It is also useful to interpret associative memories as arising from of interactions among their components. This is how Hopfield networks have been usually interpreted, where the components spins have pairwise interactions mediated by coupling constants usually defined by the Hebb rule. Modern Hopfield networks can be defined by allowing higher order interactions; this is the approach taken previously, for example, by Agliari et al.~\cite{agliari_tolerance_2020}, and also by us in this work. If we define a modern Hopfield network whose component spins interact via $n$-body interactions, with the interaction coefficients defined by the generalized Hebb rule with no noise, then its Hamiltonian will be of the form
\begin{align}
H^0_n = -\frac{1}{N^{n-1}} \sum_{1\le i_i <\cdots<i_n\le N} \sum_\mu \xi^\mu_{i_1}\cdots \xi^\mu_{i_n} \sigma_{i_1} \cdots \sigma_{i_n}, \label{intham}
\end{align}
where we have excluded self-interactions. If we had allowed self-interactions, it is clear the above Hamiltonian would have been identical to the one in eq.~\eqref{krotovham},with $F(x)=x^n$. Because $\sigma_i^2=1$, any $n$-body interaction term involving self-interactions can be written as an $(n-2k)$-body interaction term excluding self-interactions for some nonnegative integer $k \le n/2$. Thus, any Hamiltonian of the form in eq.~\eqref{krotovham}, where $F$ is an analytic function, can be written as a sum of interaction Hamiltonians of the form expressed in eq.~\eqref{intham}. In particular, by expanding the powers and using combinatorics, we see that
\begin{align}
-N \sum_\mu m_\mu &= H^0_1, \\
-N \sum_\mu m_\mu^2 &= 2H^0_2-1, \\
-N \sum_\mu m_\mu^3 &= 6H^0_3 + \qty(\frac{3}{N}-\frac{2}{N^2}) H^0_1, \\
-N \sum_\mu m_\mu^4 &= 24 H^0_4 + \qty(\frac{12}{N}-\frac{16}{N^2}) H^0_2 - \qty(\frac{3}{N}-\frac{2}{N^2}),
\end{align}
and so on. For general power $n$ the expressions get complicated, but for large $N$ we can keep the leading order terms to get the approximate expression
\begin{align}
-N \sum_\mu m_\mu^n &\approx n! \sum_{k=0}^{\lceil n/2 \rceil} \frac{1}{(2N)^k k!} H^0_{n-2k} \text{ for } N\gg 1.
\end{align}

We can invert the above to express the Hamiltonian in eq.~\eqref{intham} in terms of polynomial functions of the overlaps to obtain
\begin{align}
	H^0_1 &= -N\sum_\mu m_\mu, \\
	H^0_2 &= -\frac{1}{2} N \sum_\mu m_\mu^2 + \frac{1}{2}, \\
	H^0_3 &= -N \sum_\mu \qty( \frac{1}{6} \mu^3 - \frac{1}{2N} m_\mu ), \\
	H^0_4 &= -N \sum_\mu \qty( \frac{1}{24}m_\mu^4 - \frac{1}{4N} m_\mu^2 - \frac{1}{4N} ), \text{ etc.,}
\end{align}
and for $N \gg 1$ we have
\begin{align}
    H^0_n &= -N \sum_{\mu=1}^K \sum_{k=0}^{\lfloor n/2 \rfloor} (-1)^k \frac{m_\mu^{n-2k}}{k!(n-2k)!(2N)^k}.
\end{align}

\section{Recap of results for the classic Hopfield network}\label{descriptionforbin}
\begin{figure}
    \centering
    \includegraphics[width = 0.7\linewidth]{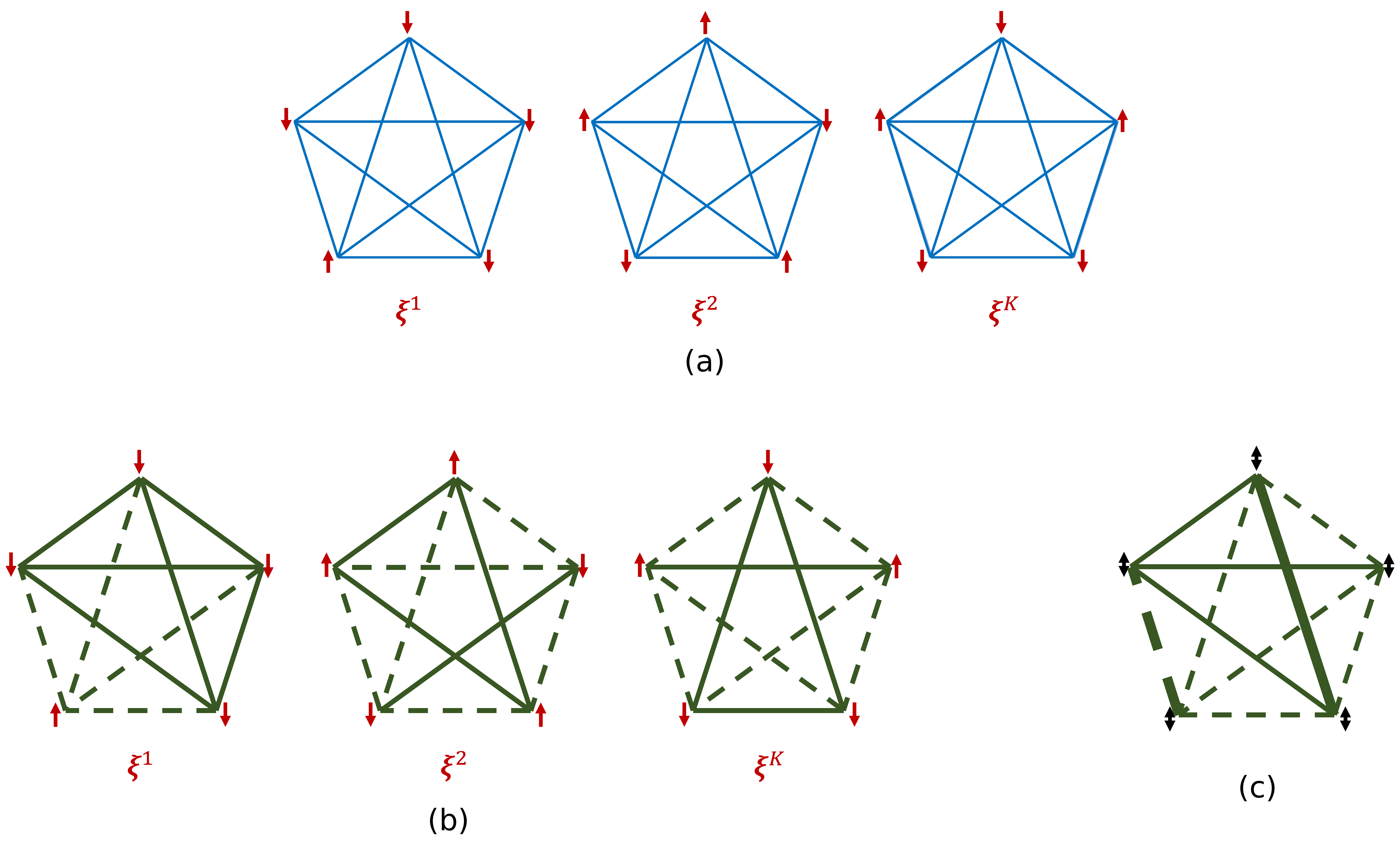}
    \caption{Schematic of a Hopfield network with $N=5$ spins and $K=3$ patterns. (a) Visual representation of the patterns, where upward and downward facing arrows correspond to a spin of $1$ and $-1$ respectively. (b) Couplings associated with each pattern (defined as $J^\mu_{ij} = \xi^\mu_i \xi^\mu_j$ for pattern $\mu$ and spins $i$ and $j$), shown as green lines connecting the arrows. Solid lines are ferromagnetic ($J^\mu_{ij}=1$) and dotted lines are antiferromganetic ($J^\mu_{ij}=-1$). (c) Couplings defined using the Hebbian learning rule as $J_{ij}=\sum_\mu J^\mu_{ij}$, shown as before by lines connecting the arrows, where the thickness of the lines indicates the strength $\abs{J_{ij}}$ of the coupling.}
    \label{fig:scheme}
\end{figure}
Here we derive the retrieval accuracy of a Hopfield network for the special case of two-body interactions (i.e., the network originally described by Hopfield~\cite{hopfield_neural_1982}). While the analysis is very similar to the general derivation shown in section~\ref{dense}, it may be helpful to study the simpler and better-known case of the standard Hopfield network and and check that our results agree with previous work. As before, we consider a system of $N$ Ising spins (binary neurons), where each spin can be in the state $1$ or $-1$, and $K$ patterns $\vb*\xi^\mu\in\qty{1,-1}^N$, $\mu\in\qty{1,\ldots,K}$, where every component $\xi^\mu_i$ independently takes the value $1$ or $-1$ with equal probability. Then the Hamiltonian for the Hopfield network is defined as an infinite-range Ising Hamiltonian with two-spin interactions,
\begin{align}
    H(\vb*\sigma) &= -\sum_{i<j} J_{ij}\sigma_i \sigma_j,
\end{align}
where the coupling constants $J_{ij}$ can be defined by the Hebb rule,
\begin{align}
    J^0_{ij} &= \sum_\mu \xi^\mu_i \xi^\mu_j.\label{binhebb}
\end{align}

Here we will generalize the coupling constants to include additive and multiplicative noise, in which case they have the form

\begin{align}
    J_{ij} &= P_{ij}\qty(\sum_\mu \xi^\mu_i \xi^\mu_j + \eta_{ij}), \label{binnoise}
\end{align}
where $P_{ij}$ is a random multiplicative factor and $\eta_{ij}$ is an additive noise term. In particular, we will mainly consider the cases where $P_{ij}$ are i.i.d.\ Bernoulli random variables with parameter $p$ (corresponding to $p$ of the couploings being retained and the rest randomly deleted), and $\eta_{ij}$ are i.i.d.\ Gaussian variables centered at $0$.

We will also consider networks where all nonzero couplings $\check J_{ij}$ have been ``clipped'' to be $\pm 1$, which we can define by taking just the sign of the coupling constant defined above, i.e.,
\begin{align}
    \check J_{ij} &= P_{ij}\sgn(\sum_\mu \xi^\mu_i \xi^\mu_j + \eta_{ij}). \label{binclip}
\end{align}

Each spin $\sigma_i$ can be thought of as experiencing a local field $B_i = \sum_{j\neq i} J_{ij} \sigma_j$, and at each time step the spins will orient themselves along their corresponding local fields. We consider synchronous updates at zero temperature, so the evolution of the system at each time step is given by the operator $\vb T$, defined such that
\begin{align}
    \qty(\vb T\vb*\sigma)_i &= \sgn(\sum_{j\neq i} J_{ij}\sigma_j) = \sgn(\sum_{\mu, j\neq i} \xi^\mu_i \xi^\mu_j\sigma_j).
\end{align}

We estimate the retrieval accuracy of the network in the manner described in section~\ref{retrieval}. We arbitrarily choose a pattern (say $\vb*{\xi}^1$) and ``corrupt'' it by randomly flipping some of its component spins with probability $\delta$. The corrupted state is $\tilde{\vb*{\xi}}^1$ with components $\tilde \xi^1_i = s_i \xi^1_i$, where $s_i$ are independent and each takes the values $1$ and $-1$ with probabilities $1-\delta$ and $\delta$ respectively. We then operate one step of the Hopfield update, and find the overlap $m = \sum_i \xi^1_i (\vb T \tilde{\vb* \xi}^1)_i/N$ of the resulting state with the original pattern to quantify how accurately the original pattern was retrieved.

Let us first consider the Hopfield network with couplings defined by the Hebb rule but with noise as in eq.~\eqref{binnoise}. Then the overlap $m$ is
\begin{align}
    m &= \frac{1}{N} \sum_i \xi^1_i \qty(
    {\vb T}\tilde{\vb* \xi}^1)_i = \frac{1}{N} \sum_i\qty(\sgn(\sum_{\mu, j\neq i} \xi^\mu_i \xi^\mu_j \xi^1_i \xi^1_j + \eta_{ij}\xi^1_i \xi^1_j) P_{ij} s_j).
\end{align}
We define the random variables $Y^\mu_{i j} = \xi^\mu_i \xi^\mu_j \xi^1_i \xi^1_j P_{ij} s_j$, $Z_{ij}=\xi^1_i\xi^1_j \eta_{ij}P_{ij}s_j$, and $X_i = \sum_{j\neq i} (\sum_\mu Y^\mu_{i j} + Z_{ij}) $. We also note that $\Expv(\xi^\mu_i)=0$ for all $\mu$ and $i$, $\Expv(s_i)=1-2\delta$ for all $i$, ${\xi^\mu_i}^2=1$ almost everywhere, $\Expv(\eta_{ij})=0$, and the variables $\qty{\xi^\mu_i}_{\mu,i}$, $\qty{\eta_{ij}}_{i<j}$, $\qty{P_{ij}}_{i<j}$, and $\qty{s_i}_i$ are all independent. We define $\mu_{\eta^2}=\Expv(\eta_{ij}^2)$, $\mu_P=\Expv(P_{ij})$, and $\mu_{P^2}=\Expv(P_{ij}^2)$. Then we can check that $\Expv(Y^\mu_{i j}) = \mu_P(1-2\delta)\delta_{\mu 1}$, $\Expv(Z_{ij})=0$, $\Varnc(Z_{ij})=\mu_{\eta^2}\mu_{P^2}$, $\Expv({X^\mu_{i j}}^2) = \mu_{P^2}-\mu_P^2(1-2\delta)^2\delta_{\mu 1}$, and all other covarances are zero. This implies $\Expv(X_i) = (N-1)\mu_P(1-2\delta)$ and $\Varnc(X_i) = (N-1) \qty((K + \mu_{\eta^2})\mu_{P^2} - \mu_P^2(1-2\delta)^2)$. Assuming that the probability distribution of $X_i$ can be approximated by a Gaussian, the probabilities of the possible values of $\sgn(X_i)$ are given by $\Proby(\sgn(X_i)=\pm 1) = \qty[1 \pm \erf(\Expv(Y_i)/\sqrt{2\Varnc(Y_i)})]/2$. Then the expected value of the overlap $m$ is
\begin{align}
    \Expv(m) = \frac{1}{N}\sum_i \Expv(\sgn(X_i)) = \erf(\sqrt{\frac{(N-1)\mu_P^2(1-2\delta)^2}{2((K+\mu_{\eta^2})\mu_{P^2}-\mu_P^2(1-2\delta)^2)}}) \approx \erf(\sqrt{\frac{N(1-2\delta)^2}{2(K+\mu_{\eta^2})\mu_{P^2}/\mu_P^2}}),\label{em2}
\end{align}
where the approximation holds when $N, K \gg 1$. We note that in absence of additive noise ($\eta_{ij}=0$), $N$ and $K$ enter this expression only through $K/N$, so any definition of the capacity involving this expected overlap will scale linearly with the number of spins. For example, we may define the capacity as the number of patterns for which the average overlap is $m_0$ for some $0<m_0 \lesssim 1$, in which case the capacity is found to be $K_\text{max} = N\mu_P^2(1-2\delta)^2/\mu_{P^2}W(2/\pi(1-m_0)^2)\approx N\mu_P^2(1-2\delta)^2/\mu_{P^2}\ln(2/\pi(1-m_0)^2)$, where $W$ is the Lambert $W$ function satisfying $W(x e^x)=x$, and $\erf^{-1}(m_0) = \sqrt{W(2/\pi(1-m_0)^2)/2}$ for $m_0$ close to $1$. This is linearly related to the number of spins $N$ as expected, and quadratically related to $1-2\delta$ quantifying how close the initial state was to the pattern being retrieved. If we assume that $P_{ij}$ are Bernoulli variables with parameter $p$, corresponding to $p$ of the couplings being retained and the rest deleted, then we have $\mu_{P}=\mu_{P^2}=p$. In that case, the capacity is also linearly related to the probability $p$ with which the couplings are retained. The capacity goes down if we discard more coupling constants or require a larger basin of attraction around each pattern, which agrees with our intuition.

\begin{figure}
    \centering
    \includegraphics[width=0.5\linewidth]{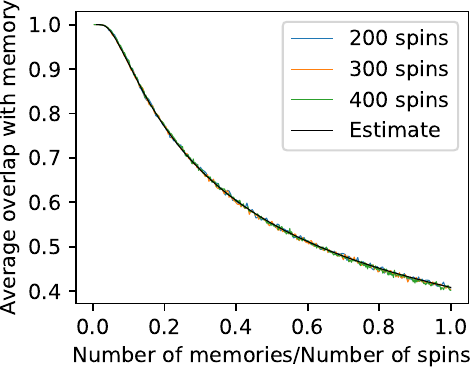}
    \caption{Average overlap of the state with the original pattern after flipping $\delta=0.2$ of the spins and doing one step of Hopfield update, in a classic Hopfield network (with $2$-spin interactions), as a function of the scaled number of patterns $\alpha = K/N$. The coupling constants are defined by the Hebb rule with no additive noise or clipping, but $p=0.8$ of the couplings have been retained and the rest deleted. The simulations were run for $N=200$, $300$, and $400$ spins, and the analytical estimate obtained in eq.~\eqref{em2} has also been plotted.}
    \label{fig:sig2u}
\end{figure}

\begin{figure}
    \centering
    \includegraphics[width=0.7\linewidth]{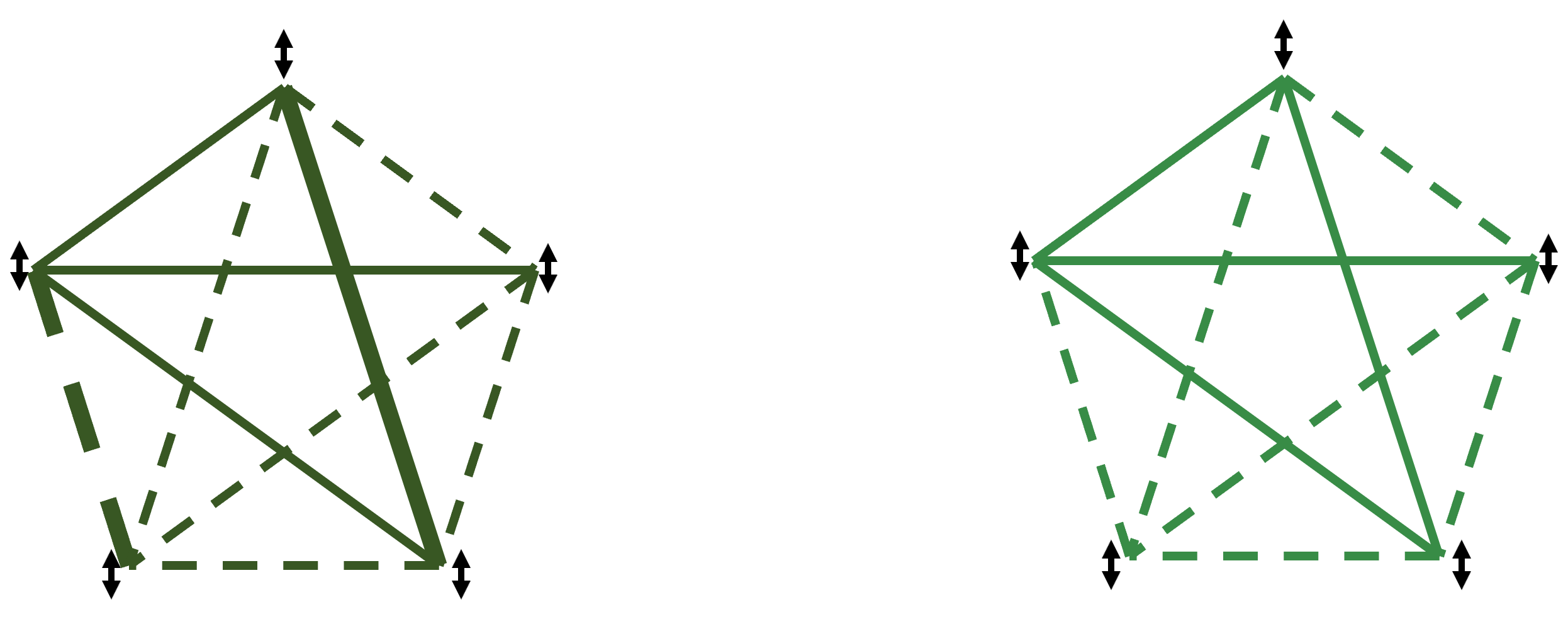}
    \caption{Schematic of a Hopfield network with $N=5$ spins and couplings restricted to be binary. The figure on the left corresponds to the model described in figure~\ref{fig:scheme}, with solid and dashed lines corresponding to ferromagnetic ($J_{ij}>0$) and antiferromagnetic ($J_{ij}<0$) interactions respectively. The figure on the right is the model with ``clipped couplings'', where all the lines have the same width (corresponding to the same magnitude of $\abs{\check J_{ij}}=1$) and can only differ in sign.}
    \label{fig:binarymodel}
\end{figure}
We now consider the model where the couplings are clipped to be $\pm 1$, following eq.~\eqref{binclip}. This ``clipped'' model has been discussed in refs.~\cite{mceliece_capacity_1987} and \cite{sompolinsky_theory_1987}; here we have also included the possibility of noise or network dilution by introducing $\eta_{ij}$ and $P_{ij}$. The overlap $\check m$ of a perturbed pattern after one step of the Hopfield update (as described in section~\ref{retrieval}) is
\begin{align}
    \check{m} &= \frac{1}{N} \sum_i\sgn(\sum_{j\neq i} \sgn(\sum_\mu \xi^\mu_i \xi^\mu_j + \eta_{ij}) \xi^1_i \xi^1_j P_{ij} s_j) = \frac{1}{N} \sum_i\sgn(\sum_{j\neq i} \sgn( \sum_\mu \xi^\mu_i \xi^\mu_j \xi^1_i \xi^1_j P_{ij} s_j ) )
\end{align}
To find the expected value of $\check m$, we define $Y^\mu_{ij} = \xi^\mu_i \xi^\mu_j \xi^1_i \xi^1_j P_{ij} s_j$ and $Z^\mu_{ij} = \xi^1_i \xi^1_j \eta_{ij} P_{ij} s_j$ as before, and also the variables $X_{ij} = \sum_\mu Y^\mu_{ij} + Z_{ij}$ and $\check X_{i} = \sum_{j\neq i}\sgn(X_{ij})$. Then for any fixed $i$, $X_i>0$ if and only if the number of variables $X_{ij}$ (for $j\neq i$) taking positive values is more than $(N-1)/2$. We see that $\Expv(X_{ij}) = \mu_{P}(1-2\delta)\equiv \bar{x}$ and $\Varnc(X_{ij}) = (K+\mu_{\eta^2})\mu_{P^2}-\mu_{P}^2(1-2\delta)^2 \equiv \sigma_x^2$. Then, assuming as before that the probability distribution is approximately Gaussian, the probability of any particular $X_{ij}$ being positive is $\Proby(X_{ij}>0) = (1+\erf(\bar{x}/\sqrt{2}\sigma_x))/2\equiv p_x$. The variables $X_{ij}$ for different $j$ are independent, so the number of positive values $N_{i} = \abs{\qty{j\neq i:X_{ij}>0}}$ follows the binomial distribution $B(N-1, p_x)$. Approximating this with with a normal distribution with the same mean and variance, we find that
\begin{align}
    \Proby(\sgn(\check X_i)=1) &= P\qty(N_i>\frac{N-1}{2}) \nn\\
    &= \frac{1}{2} + \frac{1}{2} \erf( \sqrt{ \frac{(N-1)\erf((1-2\delta)\sqrt{\mu_{P}^2/2((K+\mu_{\eta^2})\mu_{P^2}-\mu_{P}^2(1-2\delta)^2)})^2}{2-2\erf((1-2\delta)\sqrt{\mu_{P}^2/2((K+\mu_{\eta^2})\mu_{P^2}-\mu_P^2(1-2\delta)^2)})^2} } ) \nn\\
    &\approx \frac{1}{2} + \frac{1}{2} \erf((1-2\delta)\sqrt{\frac{N\mu_P^2}{\pi \mu_{P^2} (K+\mu_{\eta^2})}}),\label{clipbin}
\end{align}
where the approximation holds for $K\gg 1$. The expected value of the overlap is thus $\Expv(\check m) = \sum_i \Expv(\sgn(\check X_i))/N =\erf((1-2\delta)\mu_P\sqrt{N/\pi\mu_{P^2} (K+\mu_{\eta^2})})$, which is identical to the expression of $\Expv(m)$ that we obtained for the usual Hopfield network if we replace $N$ with $2N/\pi$. Therefore, in absence of additive noise, any measure of capacity for this clipped Hopfield network will also scale linearly with the number of spins, but will be less than the capacity of the usual Hopfield network by a factor of $2/\pi$. This agrees with the results in \cite{mceliece_capacity_1987} and \cite{sompolinsky_theory_1987}. The effects of clipping the couplings and removing them affect the capacity independently, reducing it by factors of $2/\pi$ and $p$ respectively.


\begin{figure}
    \centering
    \includegraphics[width=0.5\linewidth]{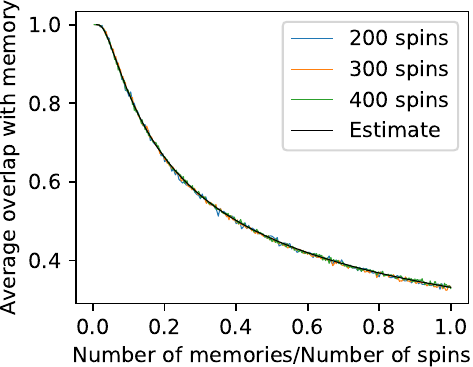}
    \caption{Average overlap of the state with the original pattern after flipping $\delta=0.2$ of the spins and doing one step of Hopfield update, in a classic Hopfield network (with 2-spin interactions), as a function of the scaled number of patterns $\alpha = K/N$. The coupling constants are defined by the Hebb rule with no additive noise, but $p=0.8$ of the spins have been clipped to have magnitude $1$ and the rest deleted entirely. The simulations were run for $N=200$, $300$, and $400$ spins, and the analytical estimate obtained in eq.~\eqref{clipbin} has also been plotted.}
    \label{fig:sig2s}
\end{figure}

\end{document}